\documentclass[12pt,a4paper]{article}
\usepackage{a4}
\usepackage{graphicx}
\usepackage{amsmath}
\usepackage{amssymb}
\usepackage{epsfig}
\usepackage{subfigure}

\newcommand{\fmn}[2]{\mbox{${\textstyle \frac{#1}{#2}}$}}
\begin{document}

\title{Review of quasi-elastic charge-exchange data
in the nucleon-deuteron breakup reaction}

\author{F.~Lehar\thanks{Email: lehar@mail.utef.cvut.cz}
\\Czech Technical University in Prague\\ IEAP, Horsk\'a
3a/22, 12800 Prague 2, Czech Republic \\[2ex]%
C.~Wilkin\thanks{Email: cw@hep.ucl.ac.uk}
\\ Physics and Astronomy Department,\\ UCL, Gower Street,
London, WC1E 6BT, UK}%
\maketitle

\begin{abstract}
The available data on the forward charge exchange of nucleons on the
deuteron up to 2~GeV per nucleon are reviewed. The value of the
inclusive $nd\to pnn/np\to pn$ cross section ratio is sensitive to
the fraction of spin-independent neutron-proton backward scattering.
The measurements of the polarisation transfer in
$d(\vec{n},\vec{p}\,)\{nn\}$ or the deuteron analysing power in
$p(\vec{d},\{pp\})n$ in high resolution experiments, where the final
$nn$ or $pp$ pair emerge at low excitation energy, depend upon the
longitudinal and transverse spin-spin $np$ amplitudes. The relation
between these types of experiments is discussed and the results
compared with predictions of the impulse approximation model in order
to see what new constraints they can bring to the neutron-proton
database.
\end{abstract}
\vspace{1cm} \noindent PACS:
{13.75.Cs}, 
{25.40.Kv}, 
{25.10.+s} 

\vspace{1cm} \noindent To be submitted to \emph{Physics of Elementary
Particles and Atomic Nuclei}

%
%
\newpage
\section{Introduction}
\label{introduction}

The charge exchange of neutrons or protons on the deuteron has a very
long history. The first theoretical papers that dealt with the
subject seem to date from the beginning of 1950s with papers by
Chew~\cite{CHE50,CHE51}, Gluckstein and Bethe~\cite{GLU51}, and
Pomeranchuk~\cite{POM51}. The first two groups were strongly
influenced by the measurements of the differential cross section of
the $d(n,p)$ reaction that were then being undertaken at UCRL by
Powell~\cite{POW51}. Apart from Coulomb effects, by charge symmetry
the cross section for this reaction should be the same as that for
$d(p,n)$. The spectrum of the emerging neutron in the forward
direction here shows a very strong peaking for an energy that is only
a little below that of the incident proton beam. There was therefore
much interest in using the reaction as a means of producing a good
quality neutron beam up to what was then ``high'' energies,
\textit{i.e.}, a few hundred MeV. The theory of this proposal was
further developed by Watson~\cite{WAT52}, Shmushkevich~\cite{SCH53},
Migdal~\cite{MIG55}, and Lapidus~\cite{LAP57}.

Since we have recently reviewed the phenomenology of the $d(n,p)$ and
$d(p,n)$ charge exchange~\cite{LEH08}, the theory will not be treated
here in any detail. The aim of the present paper is rather to discuss
the database of the existing inclusive and exclusive measurements and
make comparisons with the information that is available from
neutron-proton elastic scattering data.

The proton and neutron bound in the deuterons are in a superposition
of $^{3\!}S_1$ and $^{3\!}D_1$ states and their spins are parallel.
On the other hand, if the four-momentum transfer $t = -q^2$ between
the incident neutron and final proton in the $nd\to p\{nn\}$ reaction
is very small, the Pauli principle demands that the two emerging
neutrons be in the spin-singlet states $^{1\!}S_0$ and $^{1\!}D_2$.
In impulse (single-scattering) approximation, we would then expect
the transition amplitude to be proportional to a spin-flip
isospin-flip nucleon-nucleon scattering amplitude times a form factor
that represents the overlap of the initial spin-triplet deuteron wave
function with that of the unbound (scattering-state) $nn$ wave
function. The peaking observed in the energy spectrum of the outgoing
proton is due to the huge neutron-neutron scattering length, which
leads to a very strong final state interaction (FSI) between the two
neutrons.

A detailed evaluation of the proton spectrum from the $d(n,p)nn$
reaction would clearly depend upon the deuteron and $nn$ wave
functions, \textit{i.e.}, upon low energy nuclear physics. However, a
major advance was made by Dean~\cite{DEA72,DEA72A}. He showed that,
if one integrated over all the proton energies, there was a closure
sum rule where all the dependence on the $nn$ wave function vanished.
\begin{equation}
\label{DEAN1} \left(\frac{d\sigma}{dt}\right)_{\!\!nd\to p\{nn\}} =
        (1-F(q))\left(\frac{d\sigma}{dt}\right)_{\!\!np\to pn}^{\!\!\rm SI}
        + [1-\fmn{1}{3}F(q)]\left(\frac{d\sigma}{dt}\right)_{\!\!np\to
        pn}^{\!\!\rm SF},
\end{equation}
where $F(q)$ is the deuteron form factor. Here the neutron-proton
differential cross section is split into two parts that represent the
contribution that is independent of any spin transfer (SI) between
the initial neutron and final proton and one where there is a spin
flip (SF).

If the beam energy is high, then in the forward direction $q\approx
0$, $F(0)=1$, and Eq.~\eqref{DEAN1} reduces to
\begin{equation}
\label{DEAN2} \left(\frac{d\sigma}{dt}\right)_{\!\!nd\to p\{nn\}} =
 \frac{2}{3}\left(\frac{d\sigma}{dt}\right)_{\!\!np\to pn}^{\!\!\rm SF}.
\end{equation}
There are modifications to Eq.~\eqref{DEAN1} through the deuteron
$D$-state though these do not affect the forward limit of
Eq.~\eqref{DEAN2}~\cite{DEA72,DEA72A,BUG87}. As a consequence, the
ratio
\begin{equation}
\label{DEAN3} R_{np}(0) =
\left.\left(\frac{d\sigma}{dt}\right)_{\!\!nd\to
p\{nn\}}\right/\!\!\left(\frac{d\sigma}{dt}\right)_{\!\!np\to pn} =
\frac{2}{3}\left.\left(\frac{d\sigma}{dt}\right)_{\!\!np\to
pn}^{\!\!\rm SF}
\right/\!\!\left(\frac{d\sigma}{dt}\right)_{\!\!np\to pn}
\end{equation}
is equal to two thirds of the fraction of spin flip in $np\to pn$
between the incident neutron and proton outgoing in the beam
direction. It is because the ratio of two unpolarised cross sections
can give information about the spin dependence of neutron-proton
scattering that so many groups have made experimental studies in the
field and these are discussed in section~\ref{QECE}. Of course, for
this to be a useful interpretation of the cross section ratio the
energy has to be sufficiently high for the Dean sum rule to converge
before any phase space limitations become important. The longitudinal
momentum transfer must be negligible and terms other than the $np\to
pn$ impulse approximation should not contribute significantly to the
evaluation of the sum rule. Although the strong $NN$ FSI helps with
these concerns, all the caveats indicate that Eq.~\eqref{DEAN3} would
provide at best only a qualitative description of data at the lower
energies.

The alternative approach is not to use a sum rule but rather to
measure the excitation energy in the outgoing dineutron or diproton
with good resolution and then evaluate the impulse approximation
directly by using deuteron and $NN$ scattering wave functions,
\emph{i.e.}, input information from low energy nuclear physics. This
avoids the questions of the convergence of the sum rule and so might
yield useful results down to lower energies. A second important
feature of the $d(p,n)pp$ reaction in these conditions is that the
polarisation transfer between the initial proton and the final
neutron is expected to be very large, provided that the excitation
energy $E_{pp}$ in the final two-proton system is constrained to be
only a few MeV~\cite{PHI59,DAS68}. In fact the reaction has been used
by several groups to furnish a polarised neutron
beam~\cite{CLO80,RIL81,CHA85} but also as a means to study
neutron-proton charge exchange observables, as described in
section~\ref{transfer}.

Bugg and Wilkin~\cite{BUG87,BUG85} realised that in the small
$E_{pp}$ limit the deuteron tensor analysing powers in the
$p(\vec{d},\{pp\})n$ reaction should also be large and with a
significant angular structure that was sensitive to the differences
between the neutron-proton spin-flip amplitudes. This realisation
provided an impetus for the study of high resolution
$p(\vec{d},\{pp\})n$ experiments that are detailed in
section~\ref{polar}.

The inclusive $(p,n)$ or $(n,p)$ measurements of section~\ref{QECE}
and the high resolution ones of sections~\ref{transfer} and
\ref{polar} are in fact sensitive to exactly the same physics input.
To make this explicit, we outline in section~\ref{np_elastic} the
necessary $np$ formalism through which one can relate the forward
values of $R_{np}$ or $R_{pn}$, the polarisation transfer in
$d(\vec{n},\vec{p}\,)nn$ and the deuteron tensor analysing power in
$p(\vec{d},\{pp\})n$ in impulse approximation to the longitudinal and
transverse polarisation transfer coefficients in neutron-proton
elastic scattering. Predictions for the observables are made there
using an up-to-date phase shift analysis.

Data are available on the $R_{np}$ and $R_{pn}$ parameters in,
respectively, inclusive $d(n,p)nn$ and $d(p,n)pp$ reactions at
energies that range from tens of MeV up to 2~GeV and the features of
the individual experiments are examined in section~\ref{QECE}, where
the results are compared to the predictions of the phase shift
analysis.

Polarisation transfer data have become steadily more reliable with
time, with firmer control over the $NN$ excitation energies and
better calibrated polarisation measurements so that the data
described in section~\ref{QECE} now extend from 10~MeV up to 800~MeV.

Four experimental programmes were devoted to the study of the cross
section and tensor analysing powers of the $p(\vec{d},\{pp\})n$
reaction using very different experimental techniques. Their
procedures are described in section~\ref{polar} and the results
compared with the predictions of the plane wave impulse
approximation. In general this gives a reasonable description of the
data out to a three-momentum transfer of $q\approx m_{\pi}$ by which
point multiple scatterings might become important. These data are
however only available in an energy domain where the neutron-proton
database is extensive and reliable and the possible extensions are
also outlined there.

The comparison between the sum-rule and high resolution approaches is
one of the subjects that is addressed in our conclusions of
section~\ref{conclusions}. The consistency between the information
obtained from the $d(\vec{n},\vec{p}\,)nn$ and $p(\vec{d},\{pp\})n$
reactions in the forward direction is striking and the belief is
expressed that this must contribute positively to our knowledge of
the neutron-proton charge exchange phenomenology.

%
%
\section{Neutron-proton and nucleon-deuteron \mbox{observables}}
\label{np_elastic} \setcounter{equation}{0}

We have shown that the input necessary for the evaluation of the
forward charge exchange observables can be expressed as combinations
of \textit{pure} linearly independent $np\to np$ observables
evaluated in the backward direction~\cite{LEH08}. Although the
expressions are independent of the scattering amplitude
representation, for our purposes it is simplest to use the results of
polarisation transfer experiments. The $NN$ formalism gives two
series of polarisation transfer parameters that are mutually
dependent~\cite{BYS78}. Using the notation $X_{srbt}$ for experiments
with measured spin orientations for the scattered $(s)$, recoil
$(r)$, beam $(b)$, and target $(t)$ particles, we have either the
polarisation transfer from the beam to
recoil particles,%
\begin{equation}%
\frac{d\sigma}{dt}K_{0rb0}
=\frac{1}{4}\textit{Tr}\left\{\sigma_{2r}M\sigma_{1b}M^{\dagger}\right\},
\end{equation}
or the polarisation transfer from the target to the scattered
particle%
\begin{equation}
\frac{d\sigma}{dt}K_{s00t} =
\frac{1}{4}\textit{Tr}\left\{\sigma_{1s}M\sigma_{2t}M^{\dagger}\right\}.
\end{equation}
Here $\sigma_{1s}$, $\sigma_{1b}$, $\sigma_{2t}$, and $\sigma_{2r}$
are the corresponding Pauli matrices and $M$ is the scattering
matrix. The unpolarised invariant elastic scattering cross section
\begin{equation}
\frac{d\sigma}{dt}=\frac{\pi}{k^2}\frac{d\sigma}{d\Omega}=
\frac{1}{4}\textit{Tr}\left\{MM^{\dagger}\right\}\,,
\end{equation}
where $k$ is the momentum in the CM frame and $t$ is the
four-momentum transfer.

A first series of parameters describes the scattering of a polarised
neutron beam on an unpolarised proton target, where the polarisation
of the final outgoing protons is measured by an analyser through a
second scattering. The spins of the incident neutrons can be oriented
either perpendicularly or longitudinally with respect to the beam
direction, with the final proton polarisations being measured in the
same directions. At $\theta_{\rm CM} = \pi$ there are two independent
parameters, $K_{0nn0}(\pi)$ and $K_{0ll0}(\pi)$, referring
respectively to the transverse ($n$) and longitudinal ($l$)
directions. It was shown in Ref.~\cite{LEH08} that the forward
$d(n,p)n/p(n,p)n$ cross section ratio can be written in terms of
these as%
\begin{equation}
\label{rnp_pred} %
R_{np}(0)=\frac{1}{6}\left\{3-2K_{0nn0}(\pi)-K_{0ll0}(\pi)\right\}\,.
\end{equation}

A second series of parameters describes the scattering of an
unpolarised neutron beam on a polarised proton target, where it is
the polarisation of the final outgoing neutron that is determined.
This leads to the alternative expression for $R_{pn}(0)$:
\begin{equation}
\label{form1} R_{np}(0) = \frac{1}{6}\left\{3 - 2K_{n00n}(\pi) +
K_{l00l}(\pi)\right\},
\end{equation}
where $K_{n00n}(\pi) = K_{0nn0}(\pi)$ but $K_{l00l}(\pi) =
-K_{0ll0}(\pi)$. Other equivalent relations are to be found in
Ref.~\cite{BYS78}

It cannot be stressed enough that the small angle $(n,p)$ charge
exchange on the deuteron is sensitive to the spin transfer from the
incident neutron to the outgoing proton and NOT that to the outgoing
neutron. The latter observables are called the depolarisation
parameters $D$ which, for example, are given in the case of a
polarised target by
\begin{equation}%
\frac{d\sigma}{dt}D_{0r0t}
=\frac{1}{4}\textit{Tr}\left\{\sigma_{2r}M\sigma_{1t}M^{\dagger}\right\}.
\end{equation}
If one were to evaluate instead of Eq.~\eqref{form1} the combination
\begin{equation}
\label{form2} r_{np}(0) = \frac{1}{6}\left\{3 - 2D_{0n0n}(\pi) -
D_{0l0l}(\pi)\right\},
\end{equation}
then one would get a completely independent (and wrong) answer. Using
the SAID SP07 phase shift solution at 100~MeV one finds that
$R_{np}(0)=0.60$ while $r_{np}(0)=0.13$. Hence one has to be very
careful with the statement that the $np\to np$ spin dependence in the
backward direction is weak or strong. It depends entirely on which
particles one is discussing.

In plane wave impulse approximation, the one non-vanishing deuteron
tensor analysing power in the $p(d,\{pp\})n$ reaction in the forward
direction can be expressed in terms of the same spin-transfer
parameters, provided that the excitation energy in the $pp$ system is
very small such that it is in the $^{1\!}S_0$
state~\cite{BUG87,LEH08}:
\begin{equation}
\label{ann_pred}
A_{NN}(0)=\frac{2(K_{0ll0}(\pi)-K_{0nn0}(\pi))}{3-K_{0ll0}(\pi)-2K_{0nn0}(\pi)}\,\cdot
\end{equation}
In an attempt to minimise confusion, observables in the
nucleon-deuteron sector will be labelled with capital letters and
only carry two subscripts.

In the same approximation, the longitudinal and transverse
spin-transfer parameters in the $d(\vec{p},\vec{n})pp$ between the
initial proton and the final neutron emerging in the beam direction
are similarly given by
\begin{align}
\nonumber K_{LL}(0) &= -\left[\frac{1-3K_{0ll0}(\pi)+2K_{0nn0}(\pi)}
{3-K_{0ll0}(\pi)-2K_{0nn0}(\pi)}\right],\\
K_{NN}(0) &= -\left[\frac{1+K_{0ll0}(\pi)-2K_{0nn0}(\pi)}
{3-K_{0ll0}(\pi)-2K_{0nn0}(\pi)}\right]. \label{KKK}
\end{align}
Independent of any theoretical model, these parameters are related
by~\cite{PHI59,OHL72}
\begin{equation}
\label{identity} K_{LL}(0)+2K_{NN}(0) = -1.
\end{equation}

Equally generally, in the $^{1\!}S_0$ limit the forward longitudinal
and transverse deuteron tensor analysing powers are trivially
related;
\begin{equation}
\label{trivial1} A_{LL}(0)=-2A_{NN}(0)\,,
\end{equation}
and these are in turn connected to the spin-transfer coefficients
through~\cite{OHL72}
\begin{equation}
\label{Ohlsen} A_{LL}(0)=-(1+3K_{LL}(0))/2\qquad \textrm{or}\qquad
A_{NN}(0)=-(1+3K_{NN}(0))/2.
\end{equation}

We stress once again that, although Eqs.~(\ref{ann_pred},\ref{KKK})
are model dependent, Eqs.~(\ref{identity}), (\ref{trivial1}), and
(\ref{Ohlsen}) are exact if the final $pp$ system is in the
$^{1\!}S_0$ state.

The variation of the $np$ backward elastic cross section with energy
and the values of $R_{np}(0)$, $A_{NN}(0)$, and $K_{LL}(0)$ have been
calculated using the energy dependent GW/VPI PSA solution
SP07~\cite{ARN00} and are listed in Table~\ref{table1}. The relations
between the observables used in Refs.~\cite{ARN00} and \cite{BYS78}
are to be found in the SAID program.

\begin{table}[p]
\caption{Values of the $np$ backward differential cross section in
the CM system $d\sigma/d\Omega$, and in invariant normalisation
$d\sigma/dt$. Also shown are the forward $d(n,p)n/p(n,p)n$ ratio
$R_{np}(0)$, the longitudinal polarisation transfer parameter
$K_{LL}(0)$ in the $d(\vec{p},\vec{n})pp$ reaction, and the deuteron
analysing power $A_{NN}(0)$ in the $p(\vec{d},\{pp\})n$ reaction at
the same energy per nucleon. These have all been evaluated from the
plane wave impulse approximation using the energy dependent PSA of
Arndt \textit{et al.}, solution SP07~\cite{ARN00}. \label{table1}}
\begin{center}
\begin {small}
\begin{tabular}{|c|c|c|c|c|c|}
\hline
  $T_{n}$ & $d\sigma/d\Omega$
            & $d\sigma/dt$
            & $R_{np}(0)$
            & $K_{LL}(0)$
            & $A_{NN}(0)$             \\
  (GeV) & mb/sr& mb/(GeV/$c)^2$ &   &  &  \\
\hline
  0.010   & 78.74 & 52728 & 0.404 & -0.370 & -0.027   \\
\hline
  0.020   & 42.92 & 14371 & 0.433 & -0.273 & \phantom{-}0.045   \\
\hline
  0.030   & 29.84 & 6661 & 0.466& -0.167  & \phantom{-}0.125  \\
\hline
  0.040   & 23.56 & 3944 & 0.498 & -0.085 & \phantom{-}0.186  \\
\hline
  0.050   & 20.11 & 2693 & 0.525 & -0.030 & \phantom{-}0.227  \\
\hline
  0.060   & 18.04 & 2013 & 0.547 & \phantom{-}0.000 & \phantom{-}0.250  \\
\hline
  0.070   & 16.71 & 1599 & 0.565 & \phantom{-}0.014 & \phantom{-}0.260  \\
\hline
  0.080   & 15.81 & 1323 & 0.579 & \phantom{-}0.014 & \phantom{-}0.261  \\
\hline
  0.090   & 15.17 & 1129 & 0.591 & \phantom{-}0.006 & \phantom{-}0.255  \\
\hline
  0.100   & 14.68 & 983  & 0.600 & -0.008 & \phantom{-}0.244  \\
\hline
  0.120   & 13.98 & 780  & 0.613 & -0.048 & \phantom{-}0.214  \\
\hline
  0.150   & 13.27 & 592  & 0.627 & -0.118 & \phantom{-}0.162  \\
\hline
  0.200   & 12.46 & 417  & 0.639 & -0.231 & \phantom{-}0.077  \\
\hline
  0.250   & 11.88 & 318  & 0.645 & -0.327 & \phantom{-}0.005  \\
\hline
  0.300   & 11.45 & 255  & 0.645 & -0.405 & -0.054  \\
\hline
  0.350   & 11.19 & 214  & 0.644 & -0.472 & -0.104  \\
\hline
  0.400   & 11.02 & 184  & 0.639 & -0.530 & -0.148  \\
\hline
  0.450   & 10.88 & 162  & 0.631 & -0.582 & -0.186  \\
\hline
  0.500   & 10.62 & 142  & 0.621 & -0.630 & -0.223  \\
\hline
  0.550   & 10.10 & 123  & 0.608 & -0.678 & -0.259  \\
\hline
  0.600   & \phantom{2}9.45 & 105  & 0.596 & -0.726 & -0.295  \\
\hline
  0.650   & \phantom{2}9.07 & 93.4 & 0.588 & -0.762 & -0.321  \\
\hline
  0.700   & \phantom{2}8.96 & 85.8 & 0.586 & -0.773 & -0.330 \\
\hline
  0.750   & \phantom{2}8.95 & 79.9 & 0.588 & -0.769 & -0.327 \\
\hline
  0.800   & \phantom{2}8.93 & 74.7 & 0.592 & -0.761 & -0.321 \\
\hline
  0.850   & \phantom{2}8.98 & 69.9 & 0.596 & -0.754 & -0.315 \\
\hline
  0.900   & \phantom{2}8.81 & 65.5 & 0.601 & -0.748 & -0.311 \\
\hline
  0.950   & \phantom{2}8.73 & 61.5 & 0.605 & -0.744 & -0.308  \\
\hline
  1.000   & \phantom{2}8.65 & 57.9 & 0.609 & -0.740 & -0.305 \\
\hline
  1.050   & \phantom{2}8.57 & 54.7 & 0.613 & -0.737 & -0.303 \\
\hline
  1.100   & \phantom{2}8.50 & 51.7 & 0.616 & -0.735 & -0.302  \\
\hline
  1.150   & \phantom{2}8.44 & 49.1 & 0.620 & -0.735 & -0.301 \\
\hline
  1.200   & \phantom{2}8.40 & 46.8 & 0.623 & -0.736 & -0.302 \\
\hline
  1.250   & \phantom{2}8.38 & 44.9 & 0.626 & -0.739 & -0.304 \\
\hline
  1.300   & \phantom{2}8.39 & 43.2 & 0.629 & -0.740 & -0.308 \\
\hline
\end{tabular}
\end{small}
\end{center}
\end{table}

The GW/VPI PSA for proton-proton scattering can be used up to 3.0~GeV
but, according to the authors, the predictions are at best
qualitative above 2.5~GeV~\cite{ARN00}. Because this is an energy
dependent analysis, one cannot use the SAID program to estimate the
errors of any observable. Although the equivalent PSA for
neutron-proton scattering was carried out up to 1.3~GeV, very few
spin-dependent observables have been measured above 1.1~GeV.

Let us summarise the present status of the $np$ database at
intermediate energies. About 2000 spin-dependent $np$ elastic
scattering data points, involving 11 to 13 independent observables,
were determined at SATURNE 2 over large angular intervals mainly
between 0.8 and 1.1~GeV~\cite{ADA96,ADL99}. A comparable amount of
$np$ data in the region from 0.5 to 0.8~GeV was measured at
LAMPF~\cite{MCN96} and in the energy interval from 0.2 to 0.56~GeV at
PSI~\cite{AHM98}. The TRIUMF group also contributed significantly up
to 0.515~GeV~\cite{BUG80}.

The SATURNE 2 and the PSI data were together sufficient, not only to
implement the PSA procedure, but also to perform a direct amplitude
reconstruction at several energies and angles. It appears that the
spin-dependent data are more or less sufficient for this procedure at
the lower energies, whereas above 0.8~GeV there is a lack of $np$
differential cross section data, mainly at intermediate angles.

%
%
\section{Measurements of unpolarised quasi-elastic\\ charge-exchange observables}
\label{QECE} \setcounter{equation}{0}

\subsection{The $\boldsymbol{(n,p)}$ experiments}
\label{np}

The first measurement of the $d(n,p)$ differential cross section was
undertaken at UCRL by Powell in 1951~\cite{POW51}. These data at 90
MeV were reported by Chew~\cite{CHE51}, though only in graphical
form, and from this one deduces that $R_{np}(0) = 0.40\pm 0.04$. A
year later Cladis, Hadley, and Hess, working also at the UCRL
synchrocyclotron, published data obtained with the 270~MeV neutron
beam~\cite{CLA52}. Their value of $0.71\pm 0.02$ for the ratio of
their own deuteron/hydrogen data is clearly above the permitted limit
of 2/3 by more than the claimed error bar. This may be connected with
the very broad energy spectrum of the incident neutron beam, which
had a $\textrm{FWHM}\approx 100$~MeV.

At the Dubna synchrocyclotron the first measurements were carried out
by Dzhelepov \textit{et al.}~\cite{DZH55,DZH56} in 1952 - 1954 with a
380~MeV neutron beam. Somewhat surprisingly, the authors considered
that their result, $R_{np}(0) = 0.20\pm 0.04$, to be compatible with
the UCRL measurements~\cite{POW51,CLA52}. In fact, later more refined
experiments~\cite{PAG88} showed that the Dzhelepov \textit{et al.}\
value was far too low and it should be discarded from the database.

At the end of that decade Larsen measured the same quantity at LRL
Berkeley at the relatively high energy of 710~MeV and obtained
$R_{np}(0) = 0.48\pm 0.08$~\cite{LAR60}. However, no previous results
were mentioned in his publication.

In his contribution to the 1962 CERN conference~\cite{DZH62},
Dzhelepov presented the angular dependence of $R_{np}(\theta)$ at
200~MeV. Although he noted that the authors of the experiment were
Yu.~Kazarinov, V.~Kiselev and Yu.~Simonov, no reference was given and
we have found no publication. Reading the value from a graph, one
obtains $R_{np}(0) = 0.55 \pm 0.03$.

One advantage of working at very low energies, as was done in
Moscow~\cite{VOI65}, is that one can obtain a neutron beam from the
$^3$H$(d,n)^4$He reaction that is almost monochromatic. At 13.9~MeV
there is clearly no hope at all of fulfilling the conditions of the
Dean sum rule so that the value given in Table~\ref{table2} was
obtained with a very severe cut. Instead, the group concentrated on
the final state interaction region of the two neutrons which, in some
ways, is similar to the approach of the high resolution experiments
to be discussed in section~\ref{polar}. By comparing the data with
the $d(p,n)pp$ results of ref.~\cite{WON59}, it was possible to see
the effects of the Coulomb repulsion when the two protons were
detected in the FSI peak.

Though the value obtained by Measday~\cite{MEA66} at 152~MeV has
quite a large error bar, $R_{np}(0)=0.65 \pm 0.10$, this seems to be
mainly an overall systematic effect because the variation of the
result with angle is very smooth. These results show how
$R_{np}(\theta)$ approaches two thirds as the momentum transfer gets
large and the Pauli blocking becomes less important.

The 794~MeV measurement from LAMPF~\cite{BON78} is especially
detailed, with very fine steps in momentum transfer. Extrapolated to
$t=0$ it yields $R_{np}(0)=0.56 \pm 0.04$. However, the authors
suggest that the true value might be a little higher than this due to
the cut that they imposed upon the lowest proton momentum considered.

By far the most extensive $d(n,p)nn$ data set at medium energies was
obtained by the Freiburg group working at PSI, the results of which
are only available in the form of a diploma thesis~\cite{PAG88}.
However, the setup used by the group for neutron-proton backward
elastic scattering is described in Ref.~\cite{FRA99}. The PSI neutron
beam was produced through the interaction of an intense 589~MeV
proton beam with a thick nuclear target. This delivered pulses with
widths of less than 1~ns and bunch spacings of 20 or 60~ns. Combining
this with a time-of-flight path of 61~m allowed for a good selection
of the neutron momentum, with an average resolution of about 3\%
FWHM. Data were reported at fourteen neutron energies from 300 to
560~MeV, \emph{i.e.}, above the threshold for pion production so that
the results could be normalised using the cross section for $np\to
d\pi^0$, which was measured in parallel~\cite{FRA99}. Over this range
$R_{np}(0)$ showed very little energy dependence, with an average
value of $0.62\pm0.01$, which is quite close to the upper limit of
$2/3$.

At the JINR VBLHE Dubna a high quality quasi-monoenergetic polarised
neutron beam was extracted in 1994 from the Synchrophasotron for the
purposes of the $\Delta\sigma_L(np)$ measurements~\cite{SHA04,LEH05},
though this accelerator was stopped in 2005. Polarised deuterons are
not yet available from the JINR Nuclotron but, on the other hand,
intense unpolarised beams with very long spills could be obtained
from this machine. Since the final $\Delta\sigma_L$ set-up included a
spectrometer, the study of the energy dependence of $R_{np}(0)$ could
be extended up to 2.0~GeV through the measurement of seven
points~\cite{SHA09}. That at 550~MeV agrees very well with the
neighbouring PSI point~\cite{PAG88} while the one at 800~MeV is
consistent with the LAMPF measurement~\cite{BON78}. Since the values
of $R_{np}(0)$ above 1~GeV could not have been reliably predicted
from previous data, the Nuclotron measurements in the interval $1.0 <
T_{n}< 2$~GeV can be considered to be an important achievement in
this field. It would be worthwhile to complete these experiments by
measurements in smaller energy steps in order to recognise possible
anomalies or structures. It is also desirable to extend the
investigated interval up to the highest neutron energy at the
Nuclotron ($\approx 3.7$~GeV) since such measurements are currently
only possible at this accelerator.

The data on $R_{np}(0)$ from the $d(n,p)nn$ experiments discussed
above are summarised in Table~\ref{table2}, where the kinetic energy,
facility, year of publication, and reference are also listed. Several
original papers show the values of the angular distribution of the
charge exchange cross section on the deuteron. In such cases, the
$R_{np}(0)$ listed here were obtained using the predictions for the
free forward $np$ charge-exchange cross sections taken from the SAID
program (solution SP07)~\cite{ARN00}. These values are shown in
Table~\ref{table1}.

\begin{table}[hp]
\caption{The $R_{np}(0)$ data measured using the $d(n,p)nn$ reaction.
The total estimated uncertainties quoted do not take into account the
influence of the different possible choices on the cut on the final
proton momentum.\label{table2}}
\begin{center}
\begin{tabular}{|c|c|c|c|c|}
\hline
 &&&& \\
 $T_{n}$ & $R_{np}(0)$
                                  & Facility & Year & Ref.\\
  (MeV)    &                      &            &      &     \\
\hline
  ~13.9 & $0.19~~~~~~~~~$   & Moscow     & 1965 &\cite{VOI65}  \\
\hline
  ~90.0 & $0.40 \pm 0.04$   & UCRL       & 1951 &\cite{POW51}  \\
\hline
  152.0 & $0.65 \pm 0.10$   & Harvard    & 1966 &\cite{MEA66}  \\
\hline
  200.0 & $0.55 \pm 0.03$   & JINR DLNP  & 1962 &\cite{DZH62}  \\
\hline
  270.0 & $0.71 \pm 0.02$   & UCRL       & 1952 &\cite{CLA52}  \\
\hline
  299.7 & $0.65 \pm 0.03$   & PSI        & 1988 &\cite{PAG88}  \\
\hline
  319.8 & $0.64 \pm 0.03$   & PSI        & 1988 &\cite{PAG88}  \\
\hline
  339.7 & $0.64 \pm 0.03$   & PSI        & 1988 &\cite{PAG88}  \\
\hline
  359.6 & $0.63 \pm 0.03$   & PSI        & 1988 &\cite{PAG88}  \\
\hline
  379.6 & $0.64 \pm 0.03$   & PSI        & 1988 &\cite{PAG88}  \\
\hline
  380.0 & $0.20 \pm 0.04$   & INP Dubna  & 1955 &\cite{DZH55}  \\
\hline
  399.7 & $0.61 \pm 0.03$   & PSI        & 1988 &\cite{PAG88}  \\
\hline
  419.8 & $0.62 \pm 0.03$   & PSI        & 1988 &\cite{PAG88}  \\
\hline
  440.0 & $0.63 \pm 0.03$   & PSI        & 1988 &\cite{PAG88}  \\
\hline
  460.1 & $0.61 \pm 0.03$   & PSI        & 1988 &\cite{PAG88}  \\
\hline
  480.4 & $0.61 \pm 0.03$   & PSI        & 1988 &\cite{PAG88}  \\
\hline
  500.9 & $0.59 \pm 0.03$   & PSI        & 1988 &\cite{PAG88}  \\
\hline
  521.1 & $0.60 \pm 0.03$   & PSI        & 1988 &\cite{PAG88}  \\
\hline
  539.4 & $0.62 \pm 0.03$   & PSI        & 1988 &\cite{PAG88}  \\
\hline
  550.0 & $0.59 \pm 0.05$   & JINR VBLHE & 2009 &\cite{SHA09} \\
\hline
  557.4 & $0.63 \pm 0.03$   & PSI        & 1988 &\cite{PAG88}  \\
\hline
  710.0 & $0.48 \pm 0.08$   & LRL        & 1960 &\cite{LAR60} \\
\hline
  794.0 & $0.56 \pm 0.04$   & LAMPF      & 1978 &\cite{BON78} \\
\hline
  800.0 & $0.55 \pm 0.02$   & JINR VBLHE & 2009 &\cite{SHA09} \\
\hline
 1000~  & $0.55 \pm 0.03$   & JINR VBLHE & 2009 &\cite{SHA09} \\
\hline
 1200~  & $0.55 \pm 0.02$   & JINR VBLHE & 2009 &\cite{SHA09} \\
\hline
 1400~  & $0.58 \pm 0.04$   & JINR VBLHE & 2009 &\cite{SHA09} \\
\hline
 1800~  & $0.57 \pm 0.03$   & JINR VBLHE & 2009 &\cite{SHA09} \\
\hline
 2000 ~ & $0.56 \pm 0.05$   & JINR VBLHE & 2009 &\cite{SHA09} \\
\hline

\end{tabular}
\end{center}
\end{table}
%
%

\subsection{The $\boldsymbol{(p,n)}$ experiments}
\label{pn}

Although high quality proton beams have been available at many
facilities, the evaluation of a $R_{pn}(0)$ ratio from $d(p,n)pp$
experiments requires the division of this cross section by that for
the charge exchange on a nucleon target. Where necessary, we have
done this using the predictions of the SP07 SAID
solution~\cite{ARN00} given in Table~\ref{table1}. Given also the
difficulties in obtaining absolute normalisations when detecting
neutrons, we consider that in general the results obtained using
neutron beams are likely to be more reliable.

The low energy data of Wong \textit{et al.}~\cite{WON59} at 13.5~MeV
do show evidence of a peak for the highest momentum neutrons but this
is sitting on a background coming from other breakup mechanisms that
are probably not associated with charge exchange. The value given in
Table~\ref{table3} without an error bar is therefore purely
indicative.

In 1953 Hofmann and Strauch~\cite{HOF53}, working at the Harvard
University accelerator, published results on the interaction of
95~MeV protons with several nuclei and measured the $d(p,n)$ reaction
for the first time. An estimation of the charge-exchange ratio from
the plotted data gives $R_{pn}(0) = 0.48 \pm 0.03$.

The measurements at 30 and 50~MeV were made using the time-of-flight
facility of the Rutherford Laboratory (RHEL) Proton Linear
Accelerator~\cite{BAT66}. The neutron spectrum, especially at 30~MeV,
does not show a clear separation of the charge-exchange impulse
contribution from other mechanisms and the Dean sum rule is far from
being saturated. The same facility was used at the higher energies of
95 and 144~MeV, where the target was once again deuterated
polythene~\cite{LAN67}. This allowed the spectrum to be studied up to
a proton-proton excitation energy $E_{pp}\approx 14$~MeV when
neutrons from reactions on the carbon in the target contributed. It
was claimed that the cross sections obtained had an overall
normalisation uncertainty of about $\pm10\%$ and that the impulse
approximation could describe the data within this error bar.

The highest energy $(p,n)$ data were produced at LAMPF~\cite{BJO76},
where the charge-exchange peak was clearly separated from other
mechanisms, including pion production, and the conditions for the use
of the Dean sum rule were well satisfied. Their high value of
$R_{pn}(0)=0.66\pm0.08$ at 800~MeV would be reduced to $0.61$ if the
$np$ data of Table~\ref{table1} were used for normalisation instead
of those available in 1976.

The approach by the UCL group working at $T_p=135$~MeV at Harwell was
utterly different to the others. They used a high-pressure Wilson
cloud chamber triggered by counters, which resulted in a large
fraction of the 1740 photographs containing events~\cite{EST65}. This
led to the 1048 events of proton-deuteron collisions that were
included in the final data analysis. Instead of detecting the neutron
from the $d(p,n)pp$ reaction, the group measured both protons. In a
sense therefore the experiment is similar to that of the Dubna bubble
chamber group~\cite{GLA08}, but in inverted kinematics. Due to the
geometry of the counter selection system, the apparatus was blind to
protons that were emitted in a cone of laboratory angles $\theta_{\rm
lab}<10^{\circ}$ with energies above 6~MeV. Although the corrections
for the associate losses are model dependent, these should not affect
the neutrons emerging at small angles and the results were integrated
down to a neutron kinetic energy that was 8~MeV below the maximum
allowed. The differential cross sections were compared to the plane
wave impulse approximation calculations of Castillejo and
Singh~\cite{CAS60}.

The results from the various $d(p,n)pp$ experiments are summarised in
Table~\ref{table3}.

\begin{table}[thb]
\caption{The $R_{pn}(0)$ data measured using the $d(p,n)pp$ reaction.
The total estimated uncertainties quoted do not take into account the
influence of the different possible choices on the cut on the final
neutron momentum.\label{table3}}
\begin{center}
\begin{tabular}{|c|c|c|c|c|}
\hline
 &&&& \\
 $T_{kin}$ & $R_{pn}(0)$       & Facility    & Year & Ref.  \\
  (MeV)    &                   &               &      &       \\
\hline
  ~13.5 & $0.18~~~~~~~~~$  & Livermore   & 1959   &\cite{WON59}  \\
\hline
  ~30.1 & $0.14 \pm 0.04$  & RHEL       & 1967 &\cite{BAT66}  \\
\hline
  ~50.0 & $0.24 \pm 0.06$  & RHEL       & 1967 &\cite{BAT66}  \\
\hline
  ~95.0 & $0.48 \pm 0.03$  & Harvard    & 1953 &\cite{HOF53}  \\
\hline
  ~94.7 & $0.59 \pm 0.03$  & Harwell    & 1967 &\cite{LAN67}  \\
\hline
  135.0 & $0.65 \pm 0.15$  & Harwell    & 1965 &\cite{EST65}  \\
\hline
  143.9 & $0.60 \pm 0.06$  & Harwell    & 1967 &\cite{LAN67}  \\
\hline
  647.0 & $0.60 \pm 0.08$  & LAMPF      & 1976 &\cite{BJO76} \\
\hline
  800.0 & $0.66 \pm 0.08$  & LAMPF      & 1976 &\cite{BJO76} \\
\hline
\end{tabular}
\end{center}
\end{table}

%
%
\subsection{The unpolarised $\boldsymbol{dp\to ppn}$ reaction}
\label{Glagolev}

In principle, far more information is available if the two final
protons are measured in the deuteron charge exchange reaction and not
merely the outgoing neutron. This has been achieved by using a beam
of deuterons with momentum 3.35~GeV/$c$ incident on the Dubna
hydrogen bubble chamber. Because of the richness of the data
contained, the experiment has had a very long history with several
reanalyses~\cite{ALA75,ALA77a,ALA77b,GLA02,GLA08}.

Of the seventeen different final channels studied, the largest number
of events (over $10^5$) was associated with deuteron breakup. These
could be converted very reliably into cross sections by comparing the
sum over all channels with the known total cross section. Corrections
were made for the loss of elastic $dp$ scattering events at very
small angles. The $dp\to ppn$ events were divided into two
categories, depending upon whether it was the neutron or one of the
two protons that had the lowest momentum in the deuteron rest frame.
This identification of the charge-retention or charge-exchange
channels is expected to be subject to little ambiguity for small
momentum transfers. With this definition, the total cross section for
deuteron charge exchange was found to be $5.85\pm0.05$~mb.

The big advantage of the bubble chamber approach is that one can
check many of the assumptions that are made in the analysis. The
crucial one is, of course, the separation into the charge-exchange
and charge-retention events. In the latter case the distribution of
``spectator'' momenta $p_{\rm sp}$ falls smoothly with $p_{\rm sp}$
but in the charge-exchange sample there is a surplus of events for
$p_{\rm sp}\gtrsim 200$~MeV/$c$ that may be associated with the
virtual production of a $\Delta(1232)$ that de-excites through
$\Delta N\to pp$. Perhaps a fifth of the charge-exchange cross
section could be due to this mechanism~\cite{ALA77b} but,
fortunately, such events necessarily involve significant momentum
transfers and would not influence the extrapolation to $q=0$.

After making corrections for events that have larger opening
angles~\cite{GLA08}, the data analysis gives a value of
\begin{equation}
\label{Viktor} \left.\frac{d\sigma}{dt}(dp\to\{pp\}n)\right|_{t=0} =
\frac{2}{3}\left.\frac{d\sigma^{\rm
SF}}{dt}(dp\to\{pp\}n)\right|_{t=0} =30\pm4~\textrm{mb/(GeV}/c)^2,
\end{equation}
where $\sigma^{\rm SF}$ is the cross section corresponding to the
spin flip from the initial proton to the final neutron and the $2/3$
factor comes from the Dean sum rule. Some of the above error arises
from the estimation of the effects of the wide angle proton pairs and
in the earlier publication of the group~\cite{GLA02}, where the same
data set was treated somewhat differently, a lower value of
$25\pm3$~mb/(GeV/$c)^2$ was obtained.

The Dubna bubble chamber measurement can lead to a relatively precise
value of the average of the spin-spin amplitudes-squared. Using
Eq.~\eqref{Viktor} one obtains very similar information to that
achieved with the high resolution $dp\to\{pp\}n$ measurements to be
discussed in section~\ref{polar} and with very competitive error
bars. On the other hand, if the primary aim is to derive estimates
for the spin-independent contribution to the forward $np$
charge-exchange cross section, then it loses some of the simplicity
and directness of the $d(n,p)nn/p(n,p)n$ comparison. This is because
one has to evaluate the ratio of two independently measured numbers,
each of which has its own normalisation uncertainty. The problem is
compounded by the fact that, as we have seen from the direct $(n,p)$
measurements of $R_{np}(0)$, the contribution of the spin-independent
amplitude represents only a small fraction of the total.

In the earlier publications by the Dubna group, the necessary
normalisation denominator was taken from the elastic neutron-proton
scattering measurements of Shepard \textit{et al.}\ at the
Pennsylvania Proton Accelerator~\cite{SHE74}. These were made at
sixteen energies and over wide angular ranges. However they disagreed
strongly with all other existing $np$ data, not only in the absolute
values, but also in the shapes of angular distributions. This problem
was already apparent at low energies, starting 182~MeV. As a result,
these data have long been discarded by physicists working in the
field and they have been removed from phase shift analysis databases,
\textit{e.g.}\ from the Saclay-Geneva PSA in 1978~\cite{BYS87}.

A much more reliable $np\to pn$ data set was provided by the ER54
group of Bizard \textit{et al.}~\cite{BIZ75}, numerical values of
which are to be found in Refs.~\cite{BYS78A,BYS80}. Fitting these
data with two exponentials, gives a forward cross section of
$\left.d\sigma/dt\right|_{t=0}=54.7\pm0.2$~mb/(GeV/$c)^2$, which the
Dubna group used in their final publication~\cite{GLA08}. It is very
different from the Shepard \textit{et al.}\ result~\cite{SHE74} of
$36.5\pm1.4$~mb/(GeV/$c)^2$, which the group quoted in their earlier
work~\cite{GLA02}. This difference, together with the changed
analysis corrections, accounts for the diverse values of $R_{np}(0)$
from the same experiment that are given in Table~\ref{table4}.
%
%
\begin{table}[thb]
\caption{Summary of the available experimental data on the
$R_{np}(0)$ ratio measured with the Dubna bubble chamber using the
$dp\to \{pp\}\,n$ reaction. The kinetic energy quoted here is the
energy per nucleon. The error bars reflect both the statistical and
systematic uncertainties. Although the data sets are basically
identical, the 2008 analysis~\cite{GLA08} is believed to be the most
reliable.\label{table4}}
\begin{center}
\begin{tabular}{|c|c|c|c|c|}
\hline
 &&&& \\
 $T_{\rm kin}$ & $R_{np}(0)$      & Facility  & Year & Ref.\\
    (MeV)      &                  &             &      &       \\
\hline
  977 & $0.43\pm0.22$  & JINR VBLHE & 1975 &~\cite{ALA75} \\
  977 & $0.63\pm0.12$  & JINR VBLHE & 2002 &~\cite{GLA02} \\
  977 & $0.55\pm0.08$  & JINR VBLHE & 2008 &~\cite{GLA08} \\
\hline
\end{tabular}
\end{center}
\end{table}
%
%
\subsection{Data summary}
\label{summary}

\begin{figure}[hbt]
\centering
\includegraphics[width=0.7\columnwidth,clip]{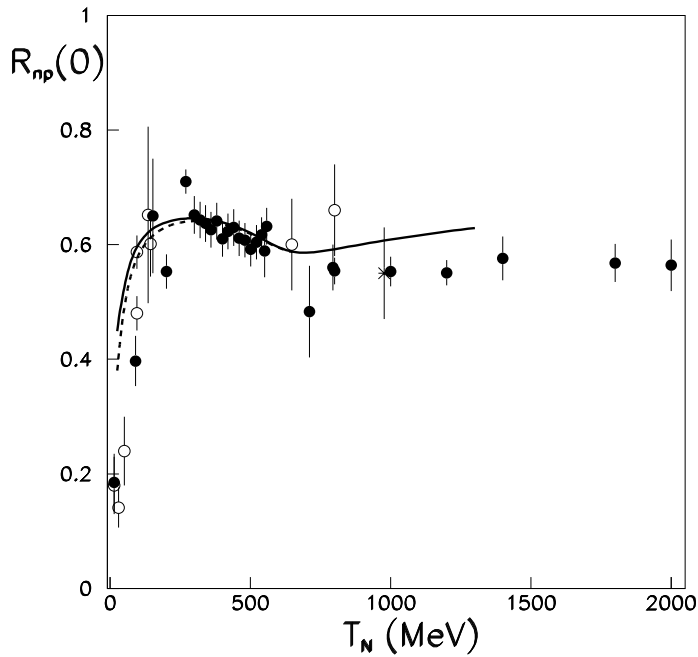}
\caption{Experimental data on the $R_{np}(0)$ ratio taken in the
forward direction. The closed circles are from the $(n,p)$ data of
Table~\ref{table2}, the open circles from the $(p,n)$ data of
Table~\ref{table3}, and the cross from the $(d,2p)$ datum of
Table~\ref{table4}. These results are compared to the predictions of
Eq.~\eqref{rnp_pred} using the current SAID solution~\cite{ARN00},
which is available up to a laboratory kinetic energy of 1.3\,GeV. The
dashed curve takes into account the limited phase space available at
the lower energies.\label{sumrule}}
\end{figure}

The values of $R_{np}(0)$ and $R_{pn}(0)$ from Tables \ref{table2}
and \ref{table3} are shown in graphical form in Fig.~\ref{sumrule},
with only the early Dubna point~\cite{DZH55} being omitted. The
$p(d,2p)$ values in Table~\ref{table4} represent the results of
increased statistics and a different analysis and only the point from
the last publication is shown~\cite{GLA08}.

The first comparison of such data with $np$ phase shift predictions
was made in 1991 in a thesis from the Freiburg group~\cite{BIN91},
where both the GW/VPI~\cite{ARN87} and Saclay-Geneva~\cite{BYS87}
were studied. The strong disagreement with the results of the PSI
measurements~\cite{PAG88} was due to the author misinterpreting the
relevant quantity as being $r_{np}(0)$ of Eq.~\eqref{form2} instead
of $R_{np}(0)$ of Eq.~\eqref{rnp_pred}.

The correct predictions from the current GW/VPI phase shift analysis
obtained on the basis of Eq.~\eqref{rnp_pred} are shown in
Fig.~\ref{sumrule} up to the limit of their validity at 1.3~GeV. The
small values of $R_{np}(0)$ at low energies is in part due to the
much greater importance of the spin-independent contribution there,
as indicated by the phase shift predictions. There are effects
arising also from the limited phase space but, when they are included
(dashed curve), they change the results only marginally. A much
greater influence is the cut that authors have to put onto the
emerging neutron or proton to try to isolate the charge-exchange
contribution from that of other mechanisms. This procedure becomes
far more ambiguous at low energies when relatively severe cuts have
to be imposed.

The data in Fig.~\ref{sumrule} seem to be largest at around the
lowest PSI point~\cite{PAG88}, where they get close to the allowed
limit of 0.67. In fact, if the Glauber shadowing effect is taken into
account~\cite{GLA67}, this limit might be reduced to perhaps 0.63. As
already shown by the phase shift analysis, the contribution from the
spin-independent term is very small in this region. On the other
hand, in the region from 1.0 to 1.3~GeV the phase shift curve lies
systematically above the experimental data. Since the conditions for
the Dean sum rule seem to be best satisfied at high energies, this
suggests that the SAID solution underestimates the spin-independent
contribution above 1~GeV. It has to be noted that the experimental
$np$ database is far less rich in this region.
%
%
\newpage
\section{Polarisation transfer measurements in
$\boldsymbol{d(\vec{p},\vec{n})pp}$}

\label{transfer}\setcounter{equation}{0}

It was first suggested by Phillips~\cite{PHI59} that the polarisation
transfer in the charge exchange reaction $d(\vec{p},\vec{n})pp$
should be large provided that the excitation energy $E_{pp}$ in the
final $pp$ system is small. Under such conditions the diproton is in
the $^{1\!}S_0$ state so that there is a spin-flip transition from a
$J^p=1^+$ to a $0^+$ configuration of the two nucleons. This
spin-selection argument is only valid for the highest neutron
momentum since, as $E_{pp}$ increases, $P$- and higher waves enter
and the polarisation signal reduces~\cite{DAS68}. Nevertheless, the
reaction has been used successfully by several groups to produce
polarised neutron beams~\cite{CLO80,RIL81,CHA85}.

In the $^{1\!}S_0$ limit, there are only two invariant amplitudes in
the forward direction and, as pointed out in Eq.~\eqref{identity},
the transverse and longitudinal spin-transfer coefficients $K_{NN}$
and $K_{LL}$ are then related by $K_{LL}(0)+2K_{NN}(0) = -1$. One
obvious experimental challenge is to get sufficient energy resolution
through the measurement of the produced neutron to guarantee that the
residual $pp$ system is in the $^{1\!}S_0$ state. The other general
problem is knowing sufficiently well the analysing power of the
reaction chosen to measure the final neutron polarisation. Some of
the earlier experiments failed on one or both of these counts.

The first measurement of $K_{NN}(0)$ for $d(\vec{p},\vec{n})pp$ seems
to have been performed at the Rochester synchrocyclotron at 200~MeV
in the mid 1960s~\cite{REA66}. A neutron polarimeter based upon $pn$
elastic scattering was used, with the analysing power being taken
from the existing nucleon-nucleon phase shifts. However, the
resolution on the final proton energies was inadequate for our
purposes, with an energy spread of 12~MeV FWHM coming from the
primary beam and the finite target thickness.

A similar experiment was undertaken at 30 and 50~MeV soon afterwards
at the RHEL Proton Linear Accelerator~\cite{ROB69}. The results
represent averages over the higher momentum part of the neutron
spectra. A liquid $^4$He scintillator was used to measure the
analysing power in neutron elastic scattering from $^4$He, though the
calibration standard was uncertain by about 8\%.

Although falling largely outside the purpose of this review, it
should be noted that there were forward angle measurements of
$K_{NN}(0)$ at the Triangle Universities Nuclear Laboratory at five
very low energies, ranging from 10.6 to 15.1~MeV~\cite{LIS80}. This
experiment also used a $^4$He polarimeter that in addition served to
measure the neutron energy with a resolution of the order of 200~keV.
Although all the data at the lowest $E_{pp}$ were consistent with
$K_{NN}(0)\approx -0.2$, a very strong dependence on the $pp$
excitation energy was found, with $K_{NN}(0)$ passing through zero in
all cases for $E_{pp}<2$~MeV. Hence, after unfolding the resolution
it is likely that the true value at $E_{pp}=0$ is probably slightly
more negative than $-0.2$. The strong variation with $E_{pp}$ is
reproduced in a simple implementation of the Faddeev equations that
was carried out, though without the inclusion of the Coulomb
interaction~\cite{JAI73}.

The RCNP experiment at 50, 65, and 80~MeV used a deuterated
polyethylene target~\cite{SAK86}. The calibration of the neutron
polarimetry was on the basis of the charge exchange from $^6$Li to
the $0^+$ ground state of $^6$Be, \textit{viz}
$^6$Li$(\vec{p},\vec{n})^6$Be$_{\rm gs}$. Although at the time the
polarisation transfer parameters for this reaction had not been
measured, they were assumed to be the same as for the transition to
the first excited (isobaric analogue) state of $^6$Li. This was
subsequently shown to be a valid assumption by a direct measurement
of neutron production with a $^6$Li target~\cite{HEN88}. On the other
hand, the resolution in $E_{pp}$ was of the order of 6~MeV, which
arose mainly from the measurement of the time of flight over 7~m. As
a consequence, the authors could not identify clearly the strong
dependence of $K_{NN}(0)$ on $E_{pp}$ that was seen in experiments
where the neutron energy was better
measured~\cite{LIS80,PIC90,ZEI99}. Such a dependence would have been
more evident in the data if there had not been a contribution at
higher $E_{pp}$ from the $^{12}$C in the target.

The most precise measurements of the polarisation transfer parameters
at low energies were accomplished in experiments at PSI at 56 and
70~MeV~\cite{PIC90,ZEI99}. One of the advantages of their setup was
the time structure of the PSI injector cyclotron, where bursts of
width 0.7~ns, separated by 20~ns, were obtained at 72~MeV, increasing
to about 1.2~ns, separated by 70~ns, at 55~MeV. This allowed the
production of a near-monoenergetic neutron beam for use in other low
energy experiments~\cite{HEN87}. Beams with a good time structure
were also obtained after acceleration of the protons to higher
energies and these were necessary for the measurements of
$R_{np}(0)$~\cite{PAG88}.

The target size was small compared to the time-of-flight path of
$\approx4.3$~m in the initial experiment~\cite{PIC90} so that the
total timing resolution of typically 1.4~ns led to one in $E_{pp}$ of
a few MeV. The polarisation of the proton beam was very well known
and that of the recoil neutron was measured by elastic scattering of
the neutrons from $^4$He. Apart from small Coulomb corrections, the
analysing power of $^4$He$(\vec{n},n)^4$He should be identical to
that of the proton in $^4$He$(\vec{p},p)^4$He, for which reliable
data existed.

The results at both 54 and 71~MeV showed that the polarisation
transfer parameters change very strongly with the measured neutron
energy and hence with $E_{pp}$. This must go a long way to explain
the anomalous results found by the RCNP group~\cite{SAK86}. At 54~MeV
both $K_{NN}$ and $K_{LL}$ were measured and, when extrapolated to
the $^{1\!}S_0$ limit of maximum neutron energy, the values gave
\begin{equation}
K_{LL}+2K_{NN} = (-0.1164\pm0.013)+2(-0.4485\pm0.011) =
-1.013\pm0.026\,,
\end{equation}
in very satisfactory agreement with the $^{1\!}S_0$ identity of
Eq.~\eqref{identity}.

The subsequent PSI measurement at 70.4~MeV made significant
refinements in two separate areas~\cite{ZEI99}. The extension of the
flight path to 11.6~m improved the resolution in the neutron energy
by about a factor of three, which allowed a much more detailed study
to the $E_{pp}$ dependence of $K_{NN}$ to be undertaken. The neutron
polarimeter used the $p(\vec{n},p)n$ reaction and an independent
calibration was carried out by studying the
$^{14}$C$(\vec{p},\vec{n})^{14}$N$_{2.31}$ reaction in the forward
direction. The 2.31~MeV level in question is the first excited state
of $^{14}$N, which is the isospin analogue of the $J^P=0^+$ ground
state of $^{14}$C. In such a case there can be no spin flip and the
polarisation of the recoil neutron must be identical to that of the
proton beam. In order to isolate this level cleanly, the neutron
flight path was increased further to 16.4~m for this target.

\begin{figure}[hbt]
\begin{center}
\includegraphics[width=0.6\columnwidth,clip]{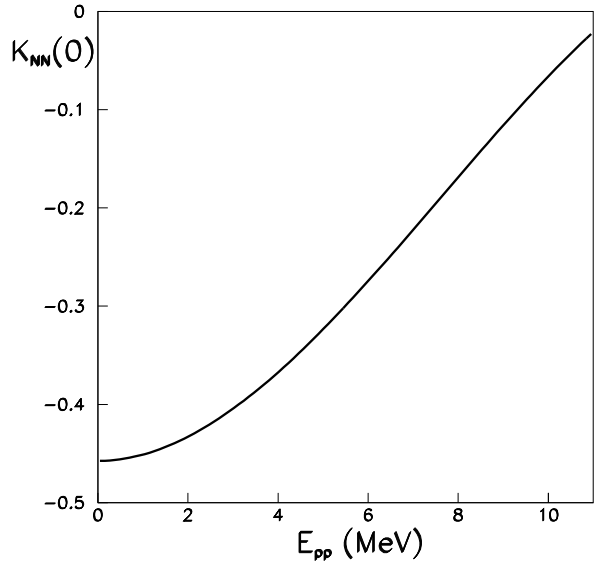}
\caption{Fit to the measured values of $K_{NN}$ of the
$d(\vec{p},\vec{n})pp$ reaction in the forward direction at a beam
energy of 70.4~MeV as a function of the excitation energy in the $pp$
final state~\cite{ZEI99}.\label{Zeier}}
\end{center}
\end{figure}

The results confirmed those of the earlier experiment~\cite{PIC90}
and, in particular, showed that even in the forward direction
$K_{NN}(0)$ varied significantly with the energy of the detected
neutron. The dependence of the parameterisation of the results on
$E_{pp}$ is shown in Fig~\ref{Zeier}. Near the allowed limit,
$E_{pp}$ is equal to the deviation of the neutron energy from its
kinematically allowed maximum.

A strong variation of the polarisation transfer parameter with
$E_{pp}$ is predicted when using the Faddeev
equations~\cite{ZEI99,GLO96}, though these do not give a perfect
description of the data. These calculations represent full multiple
scattering schemes with all binding corrections and off-shell
dependence of the nucleon-nucleon amplitudes. Nevertheless it is
important to note that the $K_{LL}(0)$ prediction for very low
$E_{pp}$ is quite close to that of the plane wave impulse
approximation. On the other hand, the fact that both the data and a
sophisticated theoretical model show the strong dependence on
$E_{pp}$ brings into question the hope that the $^{1\!}S_0$
proton-proton final state remains dominant in the forward direction
for low beam energies. This is one more reason to doubt the utility
of the Dean sum rule to estimate $R_{np}(0)$ at low energies.

The validity of the plane wave impulse approximation for the
unpolarised $d(p,n)pp$ reaction at 135~MeV has also been tested at
IUCF~\cite{AND96}. The conclusions drawn here are broadly similar to
those from an earlier study at 160~MeV~\cite{SAK87}. In the forward
direction the plane wave approach reproduces the shape of the
dependence on $E_{pp}$ out to at least 5~MeV, though the
normalisation was about 20\% too low. On the other hand, the group
evaluated the model using an $S$-state Hulth\'{en} wave function for
the deuteron and so it is not surprising that some renormalisation
was required. The $E_{pp}$ dependence follows almost exclusively from
the $pp$ wave function, which was evaluated realistically. The
comparison with more sophisticated Faddeev calculations was, of
course, hampered by the difficulty of including the Coulomb
interaction, which is particularly important for low
$E_{pp}$~\cite{PHI59}.

The values of $K_{NN}(0)$ obtained at IUCF at 160~MeV~\cite{SAK87}
show a weaker dependence on $E_{pp}$ than that found in the
experiments below 100~MeV~\cite{PIC90,ZEI99}. Nevertheless these data
do indicate that the influence of $P$-waves in the final $pp$ system
is not negligible for $E_{pp}\approx 10$~MeV.

The early measurements of $K_{NN}(0)$ and $K_{LL}(0)$ at
LAMPF~\cite{RIL81,CHA85} were hampered by the poor knowledge of the
neutron analysing power in $\vec{n}p$ elastic scattering that was
used in the polarimeter. This was noted by Bugg and
Wilkin~\cite{BUG87}, who pointed out that, although the data were
taken in the forward direction and with good resolution, they failed
badly to satisfy the identity of Eq.~\eqref{identity}. They suggested
that both polarisation transfer parameters should be renormalised by
overall factors so as to impose the condition. In view of this
argument and the results of the subsequent LAMPF
experiment~\cite{MCN92}, the values reported from these experiments
in Table~\ref{table_transfer} have been scaled such that
$K_{LL}(0)+2K_{NN}(0)=-0.98$ (to allow for some dilution from the
$P$-waves in the $pp$ system) and the error bars increased a little
to account for the uncertainty in this procedure.

The above controversy regarding the values of the forward
polarisation transfer parameters in the 500 -- 800~MeV range was
conclusively settled by a subsequent LAMPF experiment by McNaughton
\emph{et al.}\ in 1992~\cite{MCN92}. Following an idea suggested by
Bugg~\cite{BUG90}, the principle was to produce a polarised neutron
beam through the $d(\vec{p},\vec{n})pp$ reaction, sweep away the
charged particles with a bending magnet, and then let the polarised
neutron beam undergo a second charge exchange through the
$d(\vec{n},\vec{p}\,)nn$ reaction. By charge symmetry, the values of
$K_{LL}(0)$ for the two reactions are the same and, if the energy
loss in both cases is minimised, the beam polarisation $P_b$ and
final proton polarisation $P_p$ are related by
\begin{equation}
\label{McNaughton} P_p=\left[K_{LL}(0)\right]^2P_b\,.
\end{equation}
The beauty of this techniques is that only proton polarisations had
to be measured with different but similarly calibrated instruments.
Also, because the square occurs in Eq.~\eqref{McNaughton}, the errors
in the evaluation of $K_{LL}$ are reduced by a factor of two. The
energy losses were controlled by time-of-flight measurements and very
small corrections were made for the fact that the two reactions
happened at slightly different beam energies.

The overall precision achieved in this experiment was typically 3\%
and the results clearly demonstrated that there had been a
significant miscalibration in much of the earlier LAMPF neutron
polarisation standards. The group also suggested clear
renormalisations of the measured polarisation transfer parameters.
Since several of the authors of the earlier papers also signed the
McNaughton work, this lends a seal of approval to the procedure.

The longitudinal polarisation transfer in the forward direction was
measured later at LAMPF at 318 and 494~MeV~\cite{MER93} with neutron
flight paths of, respectively, 200 and 400~m so that the energy
resolution was typically 750~keV (FWHM). This allowed the authors to
use the $^{14}$C$(\vec{p},\vec{n})^{14}$N$_{2.31}$ reaction to
calibrate the neutron polarimeter, a technique that was taken up
afterwards at PSI~\cite{ZEI99}. Including these results, we now have
reliable values of either $K_{LL}(0)$ or $K_{NN}(0)$ from low
energies up to 800~MeV.

%
%

\newpage
\vspace*{-2.0cm}

\subsection{Data summary}

\begin{figure}[h!]
\begin{center}
\includegraphics[width=0.49\columnwidth,clip]{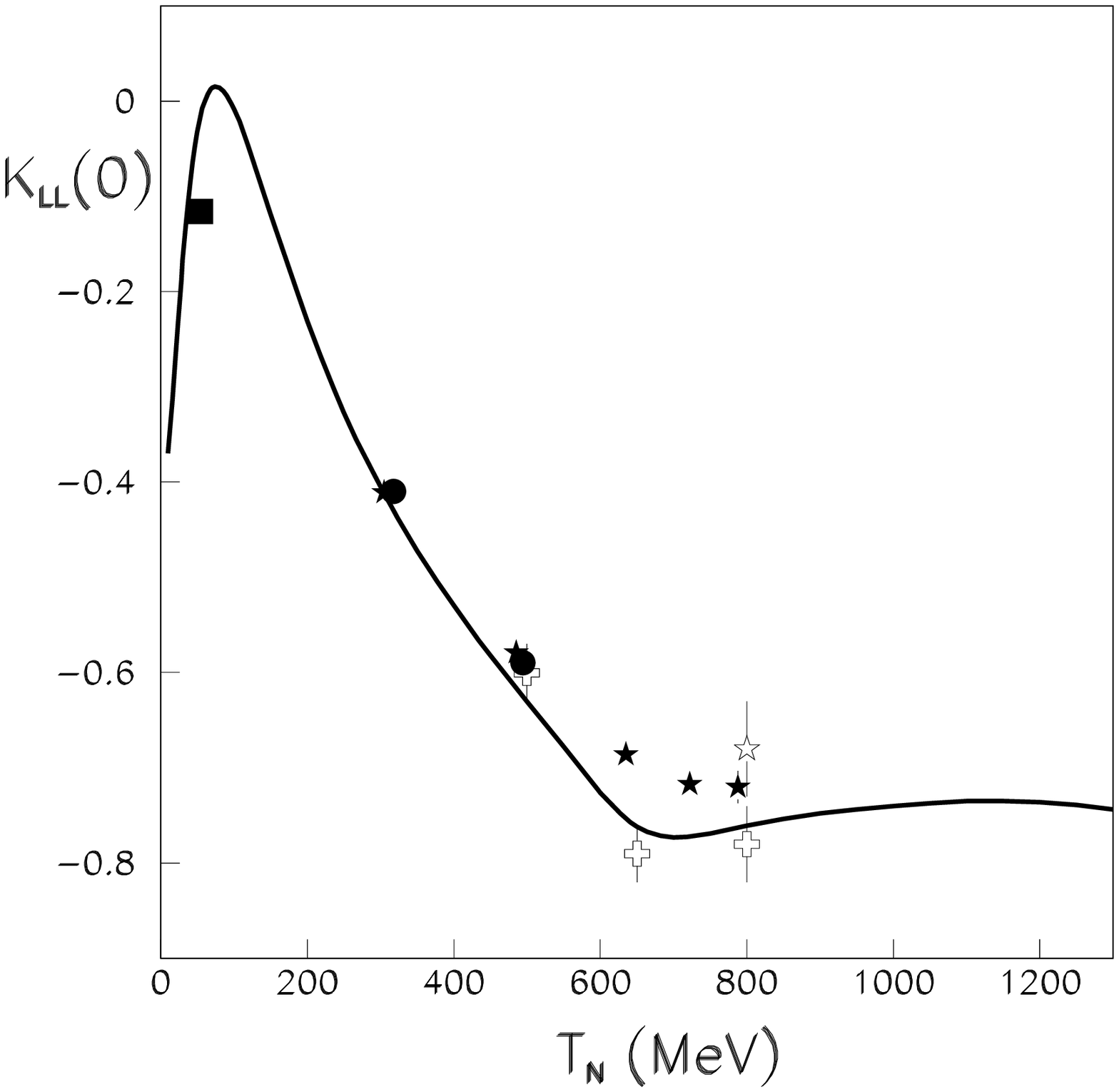}
\includegraphics[width=0.49\columnwidth,clip]{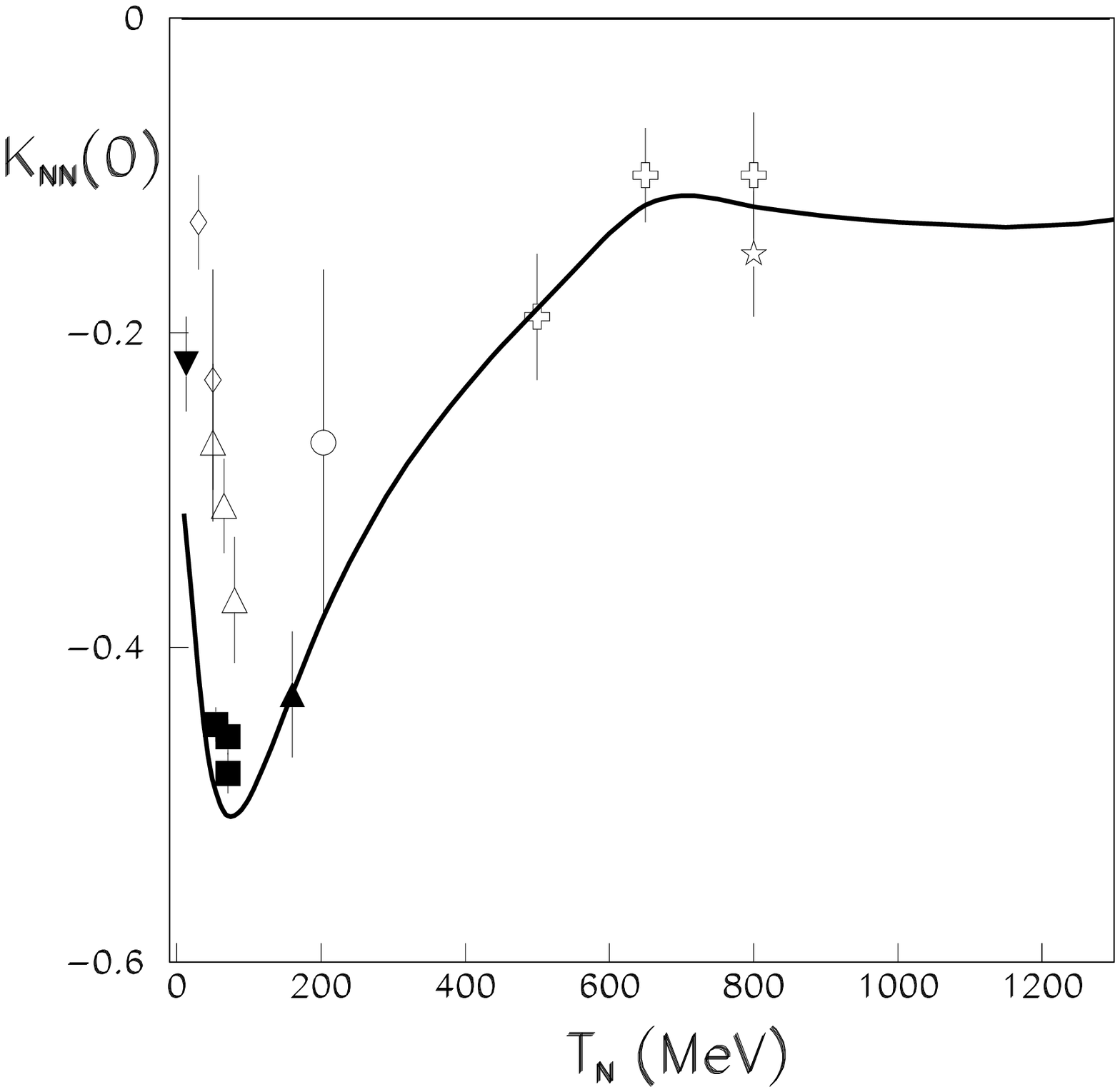}
\caption{Forward values of the longitudinal and transverse
polarisation transfer parameters $K_{LL}(0)$ and $K_{NN}(0)$ in the
$d(\vec{p},\vec{n})pp$ reaction as functions of the proton kinetic
energy $T_N$. In general we believe that greater confidence can be
placed in the data represented by closed symbols, which are from
Refs.~\cite{MCN92} (stars), \cite{MER93} (circles), \cite{SAK87}
(triangle), \cite{PIC90,ZEI99} (squares), and the average of the five
TUNL low energy points~\cite{LIS80} (inverted triangle). The open
symbols come from Refs.~\cite{ROB69} (diamonds), \cite{SAK86}
(triangle), \cite{REA66} (circle), \cite{CHA85} (crosses), and
\cite{RIL81} (star), with the latter two being renormalised as
explained in Table~\ref{table_transfer}. The curve is the plane wave
$^{1\!}S_0$ prediction of Eq.~\eqref{ann_pred}, as tabulated in
Table~\ref{table1}. \label{kllsum}}
\end{center}
\end{figure}

The values of $K_{NN}(0)$ and $K_{LL}(0)$ measured in the experiments
discussed above are presented in Table~\ref{table_transfer} and shown
graphically in Fig.~\ref{kllsum}. The results are compared in the
figure with the predictions tabulated in Table~\ref{table1} of the
pure $^{1\!}S_0$ plane wave impulse approximation of Eq.~\eqref{KKK}
that used the SAID phase shifts~\cite{ARN00} as input. Wherever
possible the data are extrapolated to $E_{pp}=0$. This is especially
important at low energies and, if this causes uncertainties or there
are doubts in the calibration standards, we have tried to indicate
such data with open symbols, leaving closed symbols for cases where
we believe the data to be more trustworthy.

The impulse approximation curve gives a semi-quantitative description
of all the data, especially the more ``reliable'' results. At low
energies we expect that this approach would be at best indicative but
it is probably significant that the curve falls below the McNaughton
\emph{et al.} results~\cite{MCN92} in the 500 to 800~MeV range, where
the approximation should be much better. It is doubtful whether the
Glauber correction~\cite{GLA67,BUG87} can make up this difference and
this suggests that the current values of the SAID neutron-proton
charge-exchange amplitudes~\cite{ARN00} might require some slight
modifications in this energy region. Similar evidence is found from
the measurements of the deuteron analysing power, to which we now
turn.

\begin{table}[hbt]
\caption{Measured values of the longitudinal and transverse
polarisation transfer parameters for the $d(\vec{p},\vec{n})pp$
reaction in the forward direction. The total estimated uncertainties
quoted do not take into account the influence of the different
possible choices on the cut on the final neutron momentum. Data
marked $^*$ have been renormalised to impose
$K_{LL}(0)+2K_{NN}(0)=-0.98$ and the error bar increased slightly.
\label{table_transfer}}
\begin{center}
\begin{small}
\begin{tabular}{|c|c|c|c|c|c|}
\hline
 &&&&& \\
 $T_{N}$ & $K_{LL}(0)$  & $K_{NN}(0)$   & Facility    & Year & Ref.  \\
  (MeV)    &              &               &             &      &       \\
\hline
  10.6  & --- & $-0.17\pm0.06$   & TUNL  & 1980   &\cite{LIS80}  \\
\hline
  12.1  & --- & $-0.20\pm0.07$   & TUNL  & 1980   &\cite{LIS80}  \\
\hline
  13.1  & --- & $-0.14\pm0.05$   & TUNL  & 1980   &\cite{LIS80}  \\
\hline
  14.1  & --- & $-0.12\pm0.06$   & TUNL  & 1980   &\cite{LIS80}  \\
\hline
  15.1  & --- & $-0.22\pm0.09$   & TUNL  & 1980   &\cite{LIS80}  \\
\hline
  ~30  & --- & $-0.13\pm0.03$   & RHEL  & 1969   &\cite{ROB69}  \\
\hline
  ~50 & ---  & $-0.23\pm0.07$   & RHEL  & 1969   &\cite{ROB69}  \\
\hline
  ~50 & ---  &$-0.27\pm0.05$   & RCNP  & 1986   &\cite{SAK86}  \\
\hline
  ~54 & $-0.116\pm0.013$ & $-0.449\pm0.011$ &PSI   & 1990   &\cite{PIC90}  \\
\hline
  ~65 & ---   &$-0.31\pm0.03$  & RCNP  & 1986   &\cite{SAK86}  \\
\hline
  70.4 & --- &$-0.457\pm0.011$& PSI   & 1999   &\cite{ZEI99}  \\
\hline
  ~71 & --- &$-0.480\pm0.013$ & PSI   & 1990   &\cite{PIC90}  \\
\hline
  ~80 & ---  & $-0.37\pm0.04$  & RCNP  & 1986   &\cite{SAK86}  \\
\hline
  160 & ---  &$-0.43\pm0.04$   & IUCF  & 1987   &\cite{SAK87}  \\
\hline
  203 & ---  &$-0.27\pm0.11$   & Rochester  & 1987   &\cite{REA66}  \\
\hline
  305 & $-0.411\pm0.010$ & --- & LAMPF & 1992   &\cite{MCN92}  \\
\hline
  318 & $-0.41\pm0.01$   & ---   & LAMPF & 1993   &\cite{MER93}  \\
\hline
  485 & $-0.579\pm0.011$ & --- & LAMPF & 1992   &\cite{MCN92}  \\
\hline
  494 & $-0.59\pm0.01$   & ---   & LAMPF & 1993   &\cite{MER93}  \\
\hline
  500 & $-0.60\pm0.03^*$ &$-0.19\pm0.04^*$ & LAMPF & 1985   &\cite{CHA85}  \\
\hline
  635 & $-0.686\pm0.012$ & --- & LAMPF & 1992   &\cite{MCN92}  \\
\hline
  650 & $-0.79\pm0.03^*$ &$-0.10\pm0.03^*$ & LAMPF & 1985   &\cite{CHA85}  \\
\hline
  722 & $-0.717\pm0.013$ & --- & LAMPF & 1992   &\cite{MCN92}  \\
\hline
  788 & $-0.720\pm0.017$ & --- & LAMPF & 1992   &\cite{MCN92}  \\
\hline
  800 & $-0.68\pm0.05^*$ &$-0.15\pm0.04^*$ & LAMPF & 1981   &\cite{RIL81}  \\
\hline
  800 & $-0.78\pm0.04^*$ &$-0.10\pm0.04^*$ & LAMPF & 1985   &\cite{CHA85}  \\
\hline
\end{tabular}
\end{small}
\end{center}
\end{table}

%
%
\clearpage
\section{Deuteron polarisation studies in high resolution $\boldsymbol{(\vec{d},2p)}$
experiments} \label{polar}\setcounter{equation}{0}

We have pointed out through Eq.~\eqref{Ohlsen} that in the
$^{1\!}S_0$ limit the deuteron $(\vec{d},2p)$ tensor analysing power
in the forward direction can be directly evaluated in terms of the
$(\vec{p},\vec{n})$ polarisation transfer coefficient. Therefore,
instead of measuring beam and recoil polarisations, much of the same
physics can be investigated by measuring the analysing power with a
polarised deuteron beam without any need to detect the polarisation
of the final particles. This is the approach advocated by Bugg and
Wilkin~\cite{BUG85,BUG87}. Unlike the sum-rule methodology applied by
a Dubna group~\cite{GLA08}, only the small part of the
$p(\vec{d},2p)n$ final phase space where $E_{pp}$ is at most a few
MeV needs to be recorded. For this purpose one does not need the
large acceptance offered by a bubble chamber and four separate groups
have undertaken major programmes using different electronic
equipment. We now discuss their results.
%
%
\subsection{The SPES~IV experiments}

The Franco-Scandinavian collaboration working at Saclay studied the
$p(\vec{d},2p)n$ reaction at 0.65, 1.6, and 2.0~GeV by detecting both
protons in the high resolution SPES~IV magnetic
spectrometer~\cite{ELL87,ELL89,SAM90,SAM95}. The small angular
acceptance ($1.7^{\circ}\times 3.4^{\circ}$) combined with a momentum
bite of $\Delta p/p \approx 7\%$ gave access only to very low $pp$
excitation energies and Monte Carlo simulations showed that the peak
of the $E_{pp}$ distribution was around 650~keV. Under these
circumstances any contamination from $P$-waves in the $pp$ system can
be safely neglected. On the other hand, the small angular acceptance
meant that away from the forward direction the data were primarily
sensitive to $A_{NN}$. On account of the small acceptance, the
deflection angle in the spectrometer was adjusted to measure the
differential cross section and $A_{NN}$ at discrete values of the
momentum transfer $q$.

\begin{figure}[hbt]
\includegraphics[width=0.49\columnwidth,clip]{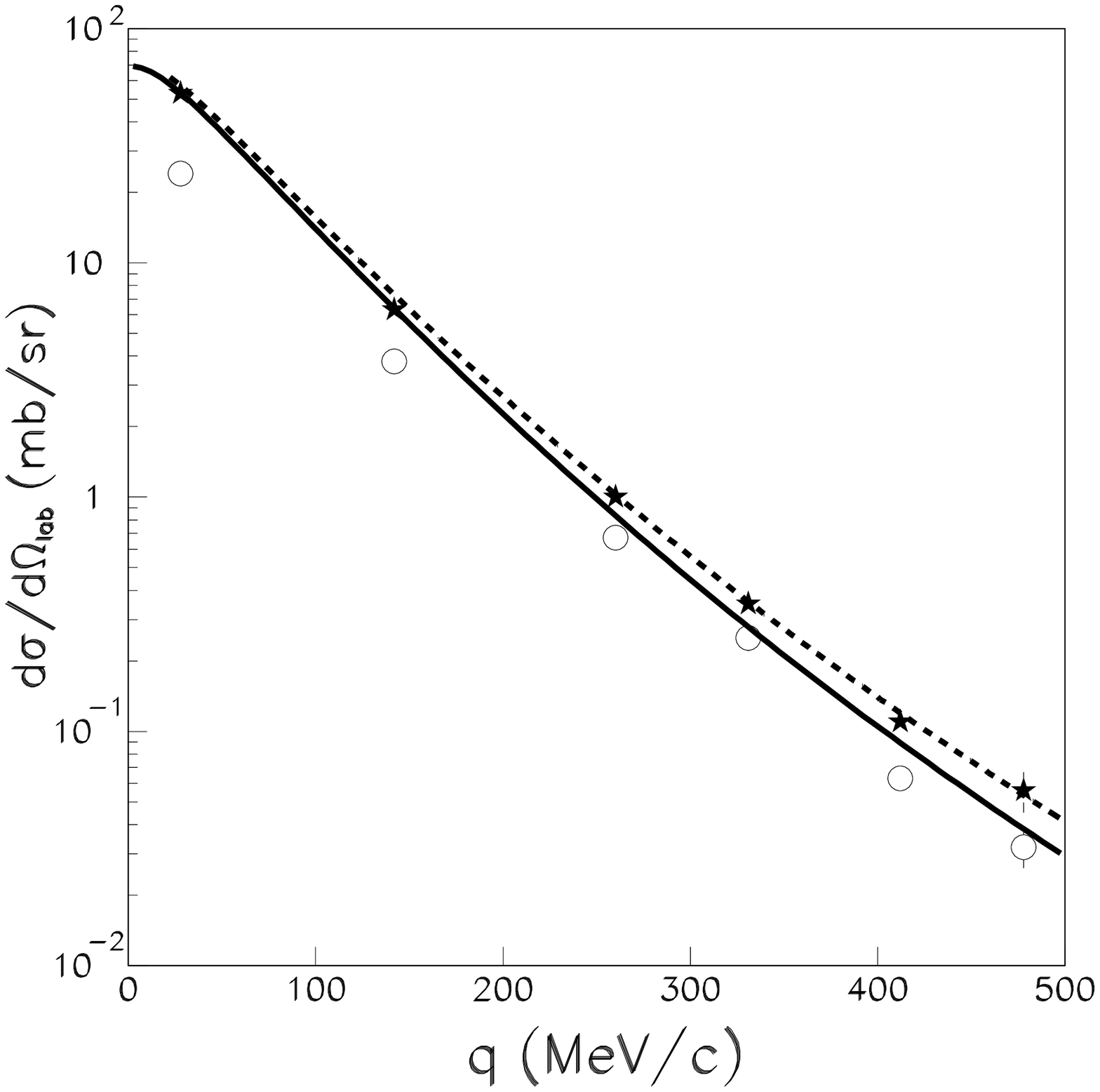}
\includegraphics[width=0.49\columnwidth,clip]{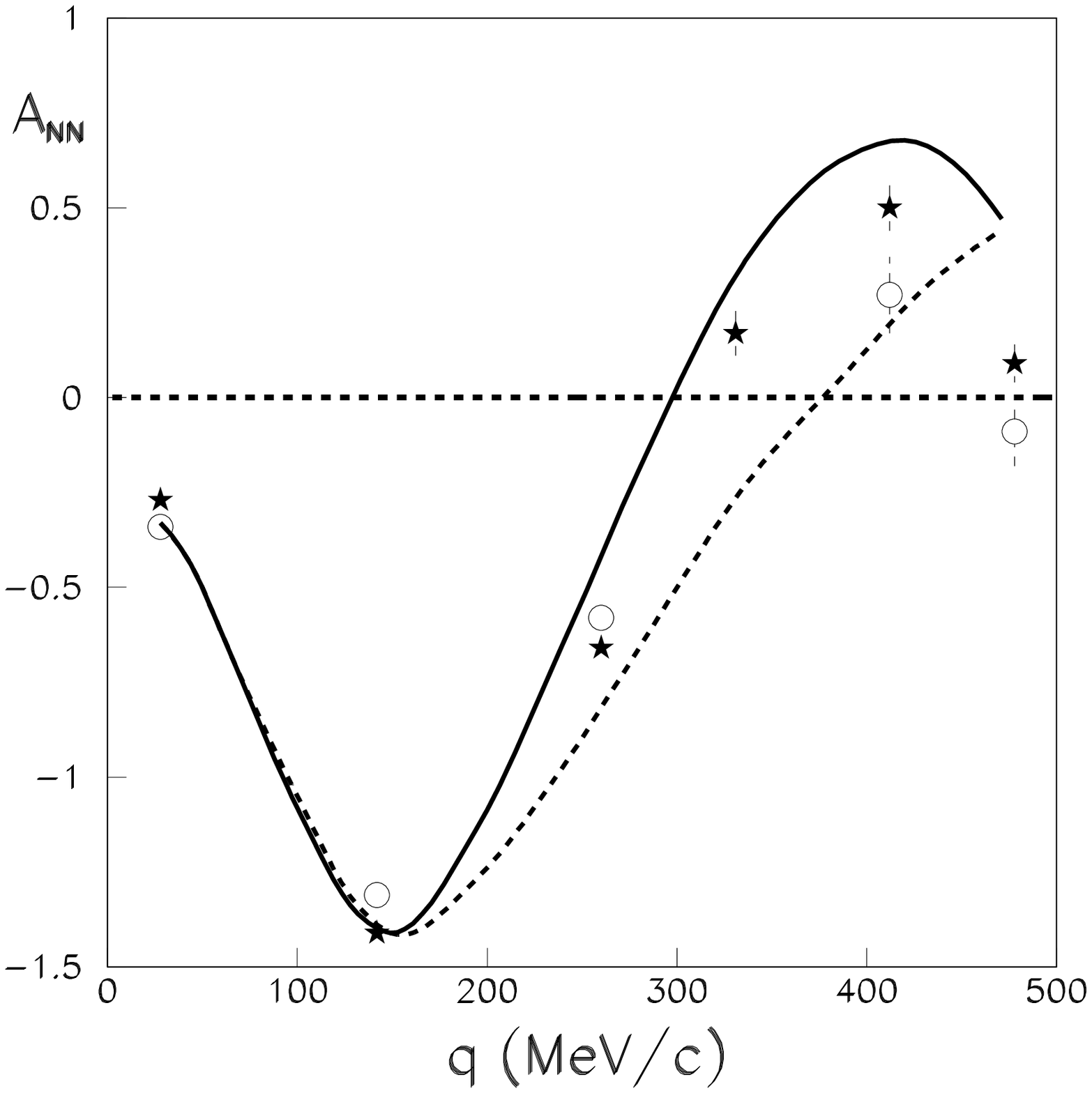}
\caption{The measurements of the $p(\vec{d},2p)n$ laboratory
differential cross section and deuteron tensor analysing power at
1.6~GeV by the Franco-Scandinavian group~\cite{SAM95} are compared to
their theoretical impulse approximation estimates without the double
scattering correction (dashed curve) and with (solid line). The
experimental cross section data (stars) have been normalised to the
solid line at $q=0.7$~fm$^{-1}$. It should be noted that the ratio of
the data on deuterium (open circles) to those on hydrogen is not
affected by this uncertainty. \label{Sams}}
\end{figure}

The results for the laboratory differential cross section and
$A_{NN}$ obtained at 1.6~GeV for both the $p(\vec{d},2p)n$ and
quasi-free $d(\vec{d},2p)nn$ reactions are shown in Fig.~\ref{Sams}.
Also shown in the figure are the authors' theoretical predictions of
the plane wave impulse approximation and also ones that included the
Glauber double-scattering term~\cite{GLA67,BUG87}. These give quite
similar results for momentum transfers below about 150~MeV/$c$ but
produce important changes for larger $q$, especially in the deuteron
analysing power. The neutron-proton charge exchange amplitudes used
were the updated versions of the analysis given in Ref.~\cite{DUB82}
that were employed in other theoretical
estimates~\cite{BUG87,CAR91,KOX91,KOX93}. The predictions were
averaged over the SPES~IV angular acceptance and, in view of the
rapid change in the transition form factor with $q$, this effect can
be significant. The validity of this procedure was tested by reducing
the horizontal acceptance by a factor of two~\cite{SAM90}.

The acceptance of the SPES~IV spectrometer for two particles was very
hard to evaluate with any precision and the hydrogen data were
normalised to the theoretical prediction at $q=0.7$~fm$^{-1}$ that
included the Glauber correction. On the other hand, the ratio of the
cross section with a deuterium and hydrogen target could be
determined absolutely and, away from the forward direction, was found
to be $0.68\pm0.04$. This is reduced even more for small $q$,
precisely because of the Pauli blocking in the unobserved $nn$
system, similar to that we discussed for the evaluation of
$R_{np}(0)$. Since for small $E_{pp}$ the $np$ spin-independent
amplitude cannot contribute and the spin-orbit term vanishes at
$q=0$, the extra reduction factor should be precisely $2/3$, which is
consistent with the value observed.

A high precision (unpolarised) $d(d,2p)nn$ experiment was undertaken
at KVI (Groningen) to investigate the neutron-neutron scattering
length~\cite{BAU05}. In this case the $pp$ and $nn$ systems were both
in the $^{1\!}S_0$ region of very small excitation energies. The
shape of the $nn$ excitation energy spectrum was consistent with that
predicted by plane wave impulse approximation with reasonable values
of the $nn$ scattering length.

The primary aim of the Franco-Scandinavian group was the
investigation of spin-longitudinal and transverse responses in medium
and heavy nuclei and also to extend these studies to the region of
$\Delta(1232)$ excitation in the $\vec{d}p\to \{pp\}\Delta^0$.
Nevertheless, it is interesting to ask how useful these data could be
for the establishment or checking of neutron-proton observables. The
$(d,2p)$ transition form factor decreases very rapidly with momentum
transfer because of the large deuteron size. As a consequence, the
Glauber double scattering term, which shares the momentum transfer
between two collisions, becomes relatively more important. Estimates
of this effect are more model dependent~\cite{BUG87,SAM95} and, as is
seen from Fig.~\ref{Sams}, it may be dangerous to rely on them beyond
about $q\approx 150$~MeV/$c$.

Absolute cross sections were not measured in these experiments and
there were only two points in the safe region of momentum transfer
and these represented averages over significant ranges in $q$. The
central values of $q$ marked on Fig.~\ref{Sams} were evaluated from a
Monte Carlo simulation of the spectrometer that used the theoretical
model as input. As a consequence, the results give relatively little
information on the magnitudes of the spin-flip compared to the
non-spin-flip amplitudes. It is perhaps salutary to note that at
larger $q$ the estimate of the cross section without the double
scattering correction describes the data better than that which
included it. However, the reverse is true for the analysing power.

The major contribution to the $np$ database comes from the
measurement of $A_{NN}$ at small $q$. Since the beam polarisation was
known with high precision, this provides a robust relation between
the magnitudes of the three spin-flip amplitudes but only at two
average values of $q$. Neutron-proton scattering has been extensively
studied in the 800~MeV region~\cite{ARN00}, and so it is not
surprising that this $p(\vec{d},2p)n$ experiment gave results that
are completely consistent with its predictions. The dip in $A_{NN}$
in both the theoretical estimates and the experimental data is due
primarily to the expected vanishing of the distorted
one-pion-exchange contribution to one of the spin-spin amplitudes for
$q\approx m_{\pi}$.
%
%
\subsection{The RCNP experiments}

Almost simultaneously with the start of the SPES~IV
experiments~\cite{ELL87}, an RCNP group studied the deuteron tensor
analysing power $A_{NN}$ in the $p(\vec{d},2p)n$ reaction at the much
lower energies of $T_d=70$~MeV~\cite{MOT88}. The primary motivation
was to compare the forward angle data with the results of the
polarisation transfer parameter $K_{NN}$ that had been measured
previously by the same group~\cite{SAK86}. For small angles a
magnetic spectrograph was used, which restricted the excitation
energy of the final protons to be less than 200~keV. At larger
angles, where the cross section is much smaller, a Si telescope array
with a larger acceptance was employed and the selection
$E_{pp}<1$~MeV was imposed in the off-line analysis. In all cases,
the only significant background arose from the random coincidence of
two protons from the breakup of separate deuterons. This is
particularly important for small angles due to the spectator momentum
distribution in the deuteron. Additional data were taken at 56~MeV,
but solely in the forward direction.

\begin{figure}[hbt]
\begin{center}
\includegraphics[width=0.7\columnwidth,clip]{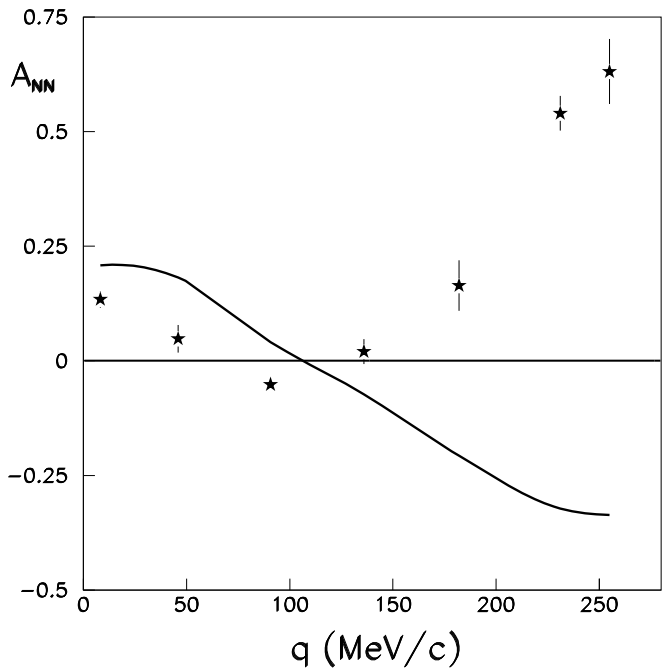}
\caption{Measurements of the deuteron tensor analysing power $A_{NN}$
for the $p(\vec{d},2p)n$ reaction at $T_d=70$~MeV by the RCNP
collaboration as a function of momentum transfer $q$~\cite{MOT88}. In
all cases $E_{pp}<1$~MeV.  The results are compared to the authors'
own theoretical plane wave impulse approximation estimates that were
based upon the SAID SP86 phase shift solution~\cite{ARN00} .
\label{Tohru}}
\end{center}
\end{figure}

At such low energies, the plane wave impulse approximation based upon
the neutron-proton charge exchange amplitudes may provide only a
semi-quantitative description of the experimental data; there are
likely to be significant contributions from direct diagrams.
Nevertheless, as can be seen in Fig.~\ref{Tohru}, the estimates given
in the paper~\cite{MOT88} that were made using the then existing
(SP86) SAID phase shift solution~\cite{ARN00} were reasonable near
the forward direction and would be even closer if modern $np$
solutions were used. At larger angles there is significant
disagreement between the data and model and the authors show that
part of this could be rectified if the $np$ input amplitudes were
evaluated at the mean of the incident and outgoing energies. This
feature has been implemented in the more refined impulse
approximation calculations of Ref.~\cite{CAR91}, where the theory was
evaluated in the brick-wall frame.

The group was disappointed to find that in the forward direction the
relation of Eq.~\eqref{Ohlsen} between their own $(\vec{p},\vec{n})$
spin transfer data~\cite{SAK86} and their deuteron tensor analysing
power was far from being satisfied. This could not be explained by
the difference in beam energy or the smearing over small angles.
Because the $(d,2p)$ results were obtained under the clean
$^{1\!}S_0$ conditions of $E_{pp}<200$~keV, the problem must be laid
at the door of the much poorer energy resolution associated with the
detection of neutrons. It was only the later PSI
experiment~\cite{ZEI99} which showed that the spin-transfer parameter
varied very strongly with $E_{pp}$ and, as argued in
section~\ref{transfer}, this is probably the resolution of the
discrepancy.
%
%

\subsection{The EMRIC experiments}

The aims and the equipment of the EMRIC
collaboration~\cite{CAR91,KOX91,KOX93}, also working at Saclay, were
very different and much closer to the original ideas of Bugg and
Wilkin~\cite{BUG85,BUG87}. The driving force was the desire to use
the $(\vec{d},2p)$ reaction as the basis for the construction of a
deuteron tensor polarimeter that could be used to measure the
polarisation of the recoil deuteron in electron-deuteron elastic
scattering. For this purpose the device had to have a much larger
acceptance than that available at SPES~IV and be compact, so that it
could be transported to and implemented in experiments at an electron
machine.

The EMRIC apparatus was composed of an array of $5 \times 5$ CsI
scintillator crystals ($4\times 4 \times 10$~cm$^3$), optically
coupled to phototubes, which provided information on both energy and
particle identification. Placed at 70~cm from the liquid hydrogen
target, it subtended an angular range of $\pm7^{\circ}$ so that
several overlapping settings were used in order to increase the
angular coverage. Since the orientation of the deuteron polarisation
could be rotated through the use of a solenoid, away from the forward
direction this gave access to both transverse deuteron tensor
analysing powers, the sideways $A_{SS}$ as well as the normal
$A_{NN}$, under identical experimental conditions.

In the initial experiment at a deuteron beam energy of
$T_d=200$~MeV~\cite{KOX91}, the angular resolution achieved with the
CsI crystals was only $\pm1.6^{\circ}$ but in the second measurement
at $T_d=350$~MeV the system was further equipped with two multiwire
proportional chambers that improved it to $0.1^{\circ}$. Having
identified fast protons using a pulse-shape analysis technique based
on the time-decay properties of the CsI crystals, their energies
could be measured with a resolution of the order of 2\%. The missing
mass of a proton pair yielded a clean neutron signal with a
$\textrm{FWHM}=14$~MeV/$c^2$, the only contamination coming from
events where not all the energy was deposited in the CsI array.

\begin{figure}[hbt]
\includegraphics[width=0.5\columnwidth,clip]{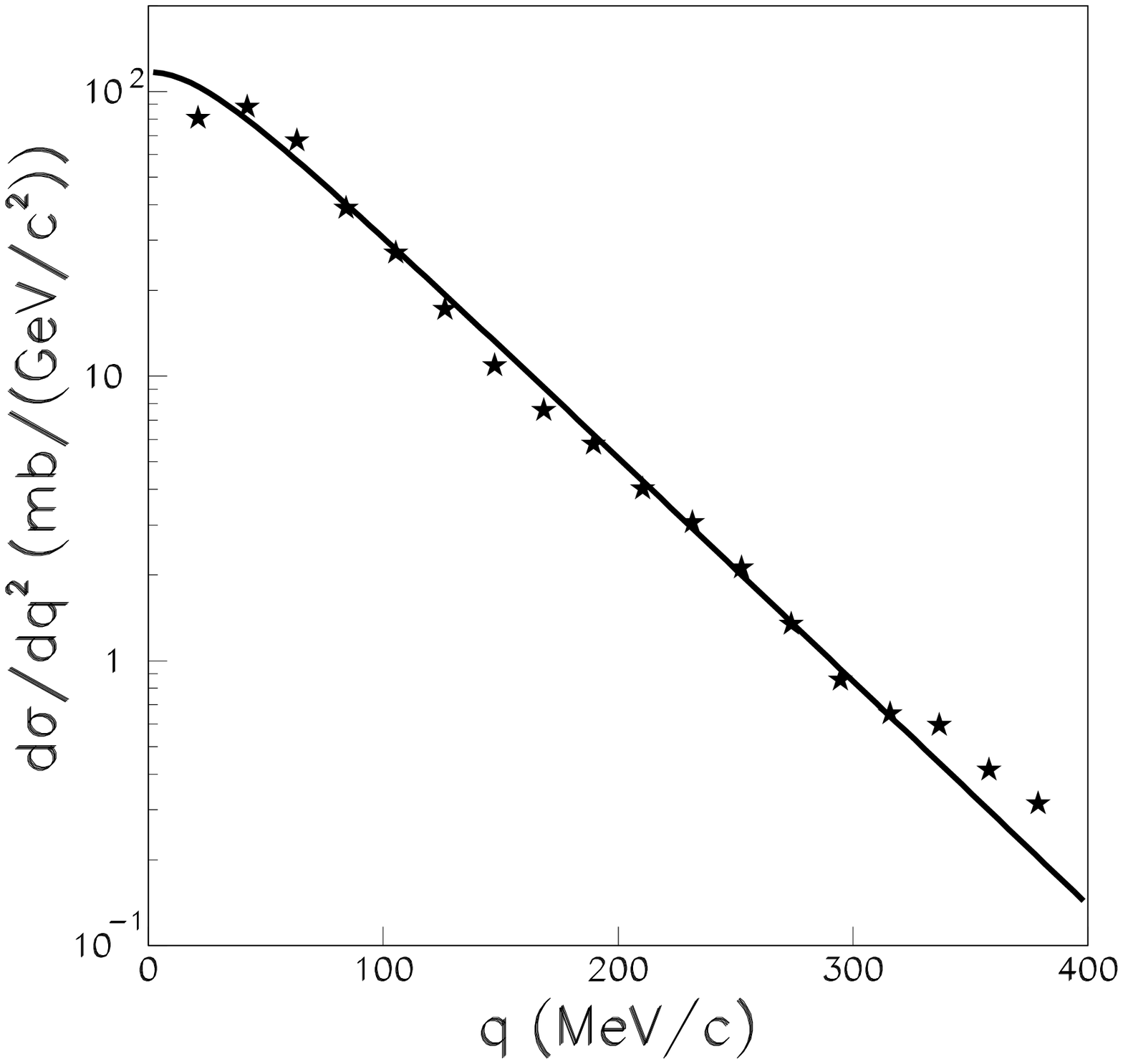}
\includegraphics[width=0.5\columnwidth,clip]{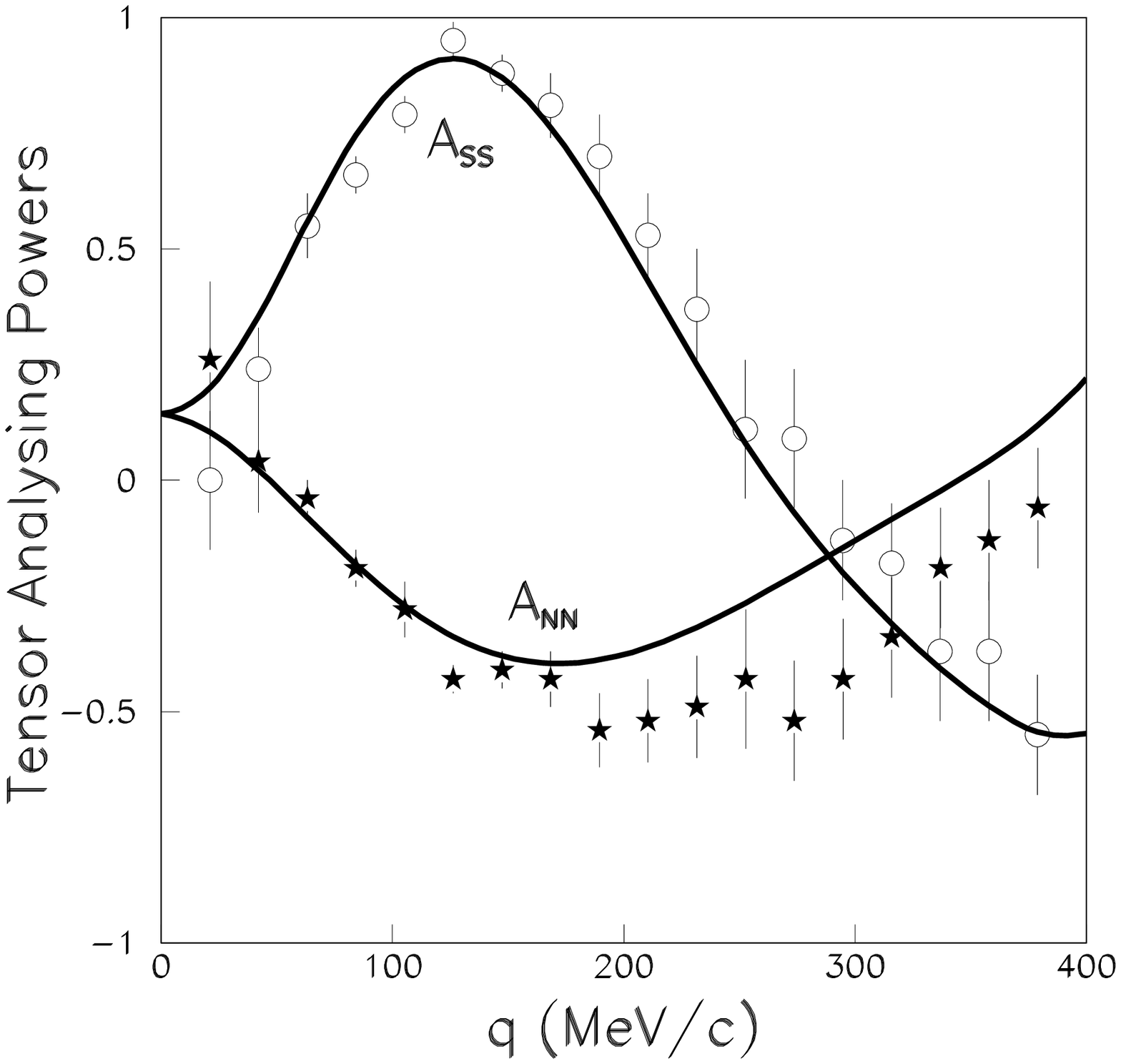}
\caption{Measurements of the $p(\vec{d},2p)n$ differential cross
section and two deuteron tensor analysing powers for $E_{pp}<1$~MeV
at a beam energy of $T_d=350$~MeV by the EMRIC
collaboration~\cite{KOX93} are compared to the theoretical plane wave
impulse approximation estimates of Ref.~\cite{CAR91}. The values of
both the experimental cross section data and theoretical model have
been scaled up by a factor of two to correct a presentational
oversight in the publication~\cite{KOX93}. \label{Serge}}
\end{figure}

The compact system allowed measurements over the wide angular and
$E_{pp}$ ranges that are necessary for the construction of a
polarimeter with a high figure of merit. However, for the present
discussion we concentrate our attention purely on the data where
$E_{pp}<1$~MeV, for which the dilution of the analysing power signal
by the proton-proton $P$ waves is small. The EMRIC results for the
differential cross section and two tensor analysing powers at 350~MeV
are shown in Fig.~\ref{Serge}. Due to a slip in the preparation of
the publication~\cite{KOX93}, both the experimental data and the
impulse approximation model were downscaled by a factor of
two~\cite{REA94}, which has been corrected in the figure shown here.
One should take into account that there are systematic errors (not
shown) arising from the efficiency corrections that are estimated to
be typically of the order of 20\%, though they are larger at the
edges of EMRIC~\cite{KOX09}. This might account for the slight
oscillations of the data around the theoretical prediction in
Fig.~\ref{Serge}.

The plane wave impulse approximation calculation of Ref.~\cite{CAR91}
describes the data quite well, though one has to note that the
presentation is on a logarithmic scale and that there are at least
20\% normalisation uncertainties. The data represented three settings
of the EMRIC facility and their fluctuations around the predictions
could be partially due to minor imperfections in the acceptance
corrections. The model is also satisfactory for the analysing powers
out to at least $q\approx 150$~MeV/$c$, from which point the $A_{NN}$
data remain too negative. However, as we argued with the SPES~IV
results of Fig.~\ref{Sams}, it is at about this value of $q$ that the
Glauber double scattering correction becomes significant. We can
therefore conclude that the good agreement of the $A_{SS}$ and
$A_{NN}$ data in the ``safe'' region of $q\lesssim 150$~MeV/$c$ is
confirmation that the ratios of the different spin-spin contributions
given by the Bugg amplitudes of Ref.~\cite{DUB82} are quite accurate.
Nevertheless, their overall strength is checked far less seriously by
these data because of the normalisation uncertainty and the
logarithmic scale of Fig.~\ref{Sams}.

The EMRIC experiment~\cite{KOX93} was the only one of those discussed
that was capable of investigating the variation of the deuteron
analysing power $A_{NN}$ with excitation energy and, in view of the
strong effects found for the $d(\vec{p},\vec{n})pp$ polarisation
transfer parameters at 56 and 70~MeV~\cite{PIC90,ZEI99}, it would be
interesting to see if anything similar happened for $A_{NN}$.
Extrapolating the $T_d=200$~MeV results to the forward direction, it
is seen that $A_{NN}\approx 0.23$, 0.17 and 0.10 for the three bins
of excitation energy $E_{pp}<1$~MeV, $1<E_{pp}<4$~MeV, and
$4<E_{pp}<8$~MeV, respectively. This variation is smaller than that
found for $K_{NN}$~\cite{PIC90,ZEI99}. On the other hand, since the
(longitudinal) momentum transfer remains very small in the forward
direction, the plane wave impulse approximation predicts very little
change with $E_{pp}$.

The aim of the group was to show that the $(\vec{d},2p)$ reaction had
a large and well understood polarisation signal and this was
successfully achieved. The experience gained with the EMRIC device
laid the foundations for the development of the POLDER
polarimeter~\cite{KOX94,REA94}, which was subsequently used to
separate the contributions from the deuteron monopole and quadrupole
form factors at JLab~\cite{ABB00}.\vspace{2cm}
%
%
\subsection{The ANKE experiments}

A fourth experimental approach is currently being undertaken using
the ANKE magnetic spectrometer that is located at an internal target
position forming a chicane in the COSY COoler SYnchrotron. This
machine is capable of accelerating and storing protons and deuterons
with momenta up to 3.7~GeV/$c$, \textit{i.e.}, kinetic energies of
$T_p=2.9$~GeV and $T_d=2.3$~GeV. The $(\vec{d},2p)$ measurements form
part of a much larger spin programme that will use combinations of
polarised beams and targets~\cite{SPI05}. Only results from a test
experiment at $T_d=1170$~MeV are presently
available~\cite{CHI06a,CHI09}, and these are described below.

There are several problems to be overcome before the $p(\vec{d},2p)n$
reaction could be measured successfully at ANKE. The horizontal
acceptance for the reaction is limited to laboratory angles in the
range of approximately $-2^{\circ}<\theta_{\textrm{hor}}<4^{\circ}$
and much less in the vertical direction. This constrains severely the
range of momentum transfers that can be studied. Furthermore, the
axis of the spin alignment of the circulating beam is vertical and,
unlike the EMRIC case~\cite{KOX93}, there is insufficient place for a
solenoid to rotate the polarisation. As a consequence, the values of
$A_{NN}$ and $A_{SS}$ cannot be extracted under identical condition.
Furthermore, the polarisations of the beam have to be checked
independently at the ANKE energy. Finally, unlike the external beam
experiments of SPES~IV or EMRIC, the luminosity inside the storage
ring has also to be established at the ANKE position.

Most of the above difficulties can be addressed by using the fact
that one can observe and measure simultaneously in ANKE the following
reactions: $\vec{d}p\to\{pp\}n$, $\vec{d}p \to dp$,
$\vec{d}p\to\,^3$He$\pi^0$, and $\vec{d}p\to p_{\rm sp}d\pi^0$, where
$p_{\rm sp}$ is a fast spectator proton. What cannot, of course, be
avoided is the cut in the momentum transfer which at $T_d=1170$~MeV
means that the deuteron charge exchange reaction has good acceptance
only for $q\lesssim 150$~MeV/$c$. However, we already saw in the
SPES~IV case that for larger momentum transfers the double scattering
corrections become important and, as a result, the extraction of
information on $np$ amplitudes becomes far more model dependent.

The luminosity, and hence the cross section, was obtained from the
measurement of the $dp\to p_{\rm sp}d\pi^0$ reaction, for which the
final spectator proton and produced deuteron fall in very similar
places in the ANKE forward detector to the two protons from the
charge exchange reaction. Using only events with small spectator
momenta, and interpreting the reaction as being due to that induced
by the neutron in the beam deuteron, $np\to d\pi^0$, reliable values
could be obtained for the luminosity. This approach had the
subsidiary advantage that to some extent the Glauber shadowing
correction~\cite{GLA67} cancels out between the $dp\to p_{\rm
sp}d\pi^0$ and $dp\to\{pp\}n$ reactions.

The COSY polarised ion source that feeds the circulating beam was
programmed to provide a sequence of one unpolarised state, followed
by seven combinations of deuteron vector and tensor polarisations.
Although these were measured at low energies, it had to be confirmed
that there was no loss of polarisation through the acceleration up to
$T_d=1170$~MeV. This was done by measuring the analysing powers of
$\vec{d}p \to dp$, $\vec{d}p\to\,^3$He$\pi^0$, and $\vec{d}p\to
p_{\rm sp}d\pi^0$ and comparing with results given in the
literature~\cite{CHI06b}. As expected, there was no discernable
depolarisation.

Due to the geometric limitations, the acceptance of the ANKE forward
detector varies drastically with the azimuthal production angle
$\phi$. The separation between $A_{NN}$ and $A_{SS}$ depends upon
studying the variation of the cross section with $\phi$. An accurate
knowledge of the acceptance is not required for this purpose because
one can work with the ratio of the polarised to unpolarised cross
section where, to first order, the acceptance effects drop out. The
Monte Carlo simulation of the acceptance was sufficiently good to
give only a minor contribution to the error in the unpolarised cross
section itself. The claimed overall cross section uncertainty of 6\%
is dominated by that in the luminosity evaluation.

\begin{figure}[hbt]
\includegraphics[width=0.5\columnwidth,clip]{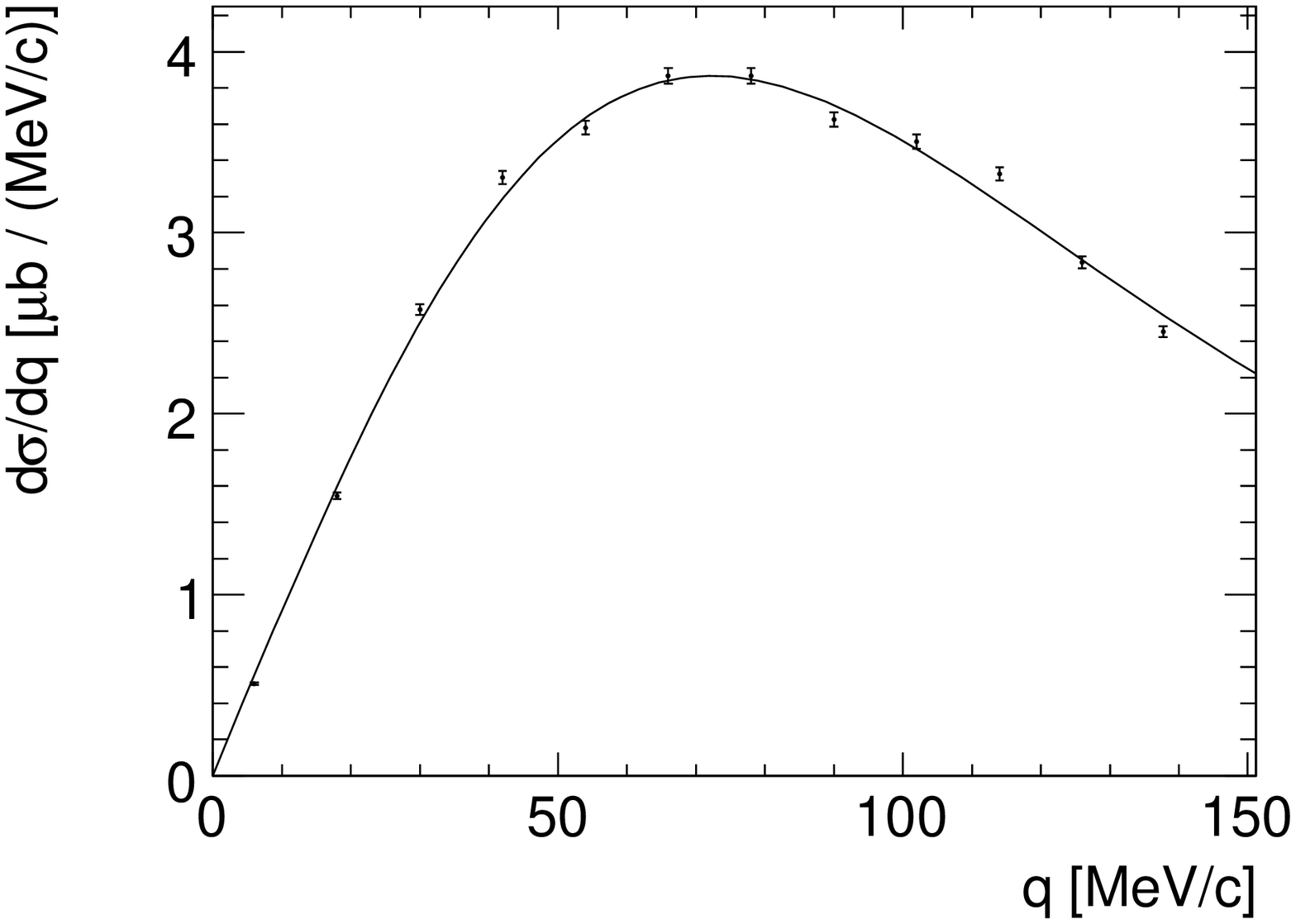}
\includegraphics[width=0.5\columnwidth,clip]{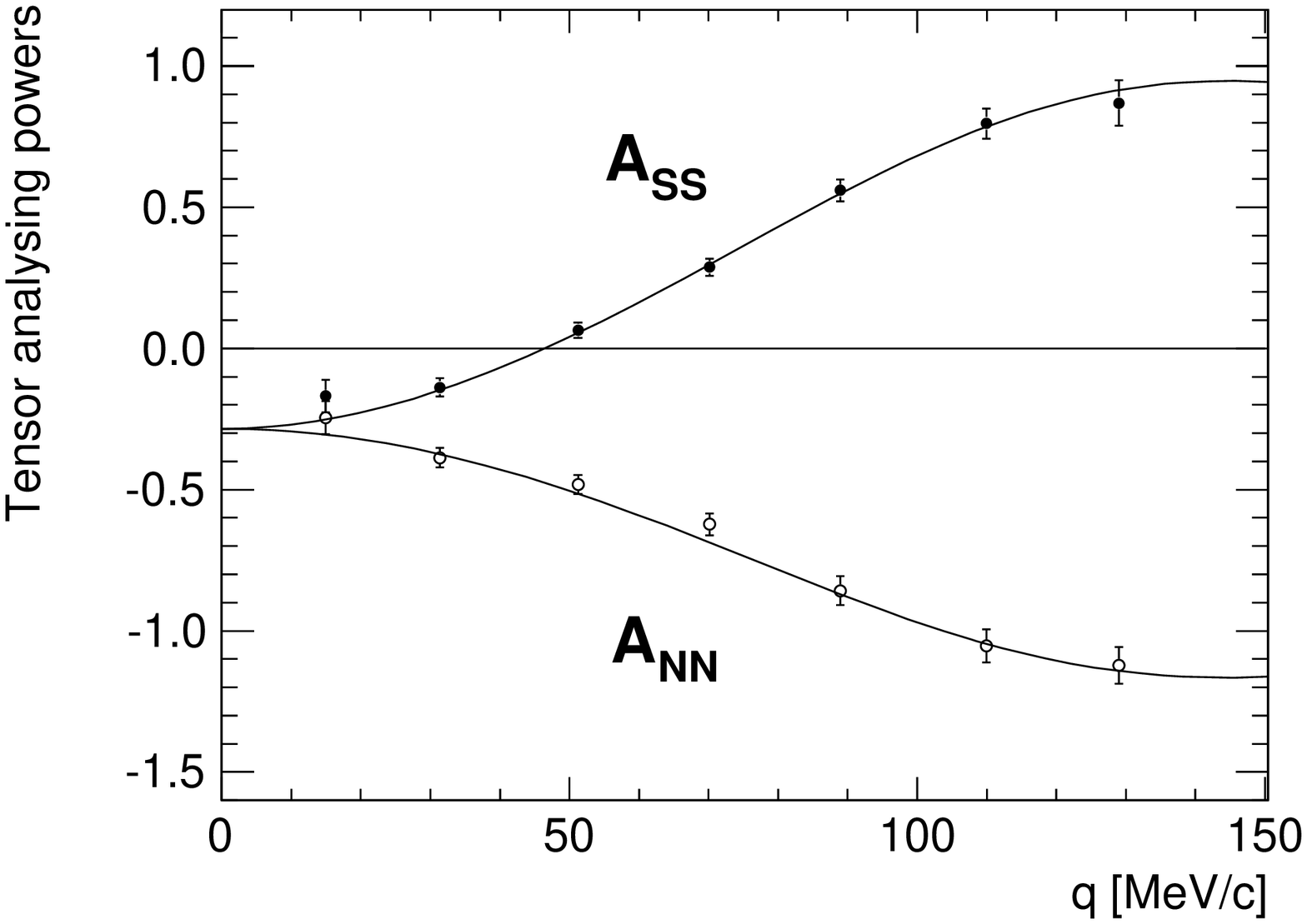}
\caption{Measurements of the $p(\vec{d},2p)n$ differential cross
section and two deuteron tensor analysing powers for $E_{pp}<1$~MeV
at a beam energy of $T_d=1170$~MeV by the ANKE
collaboration~\cite{CHI06a,CHI09} are compared to the theoretical
plane wave impulse approximation estimates of
Ref.~\cite{CAR91}.\label{David}}
\end{figure}

The limited ANKE acceptance also cuts into the $E_{pp}$ spectrum and
the collaboration only quote data integrated up to a maximum of
3~MeV. The results shown in Fig.~\ref{David} were obtained with a cut
of $E_{pp}<1$~MeV, as were the updated theoretical predictions from
Ref.~\cite{CAR91}, where the current SAID $np$ elastic amplitudes at
585~MeV were used as input~\cite{ARN00}.

The agreement between the plane wave impulse approximation and the
experimental data is very good for all three observables over the
full momentum transfer range that is accessible at ANKE. Since there
have been many neutron-proton experiments in this region, it is to be
believed that the $np$ elastic scattering amplitudes are very
reliable at 585~MeV. Extrapolating the results to $q=0$ and using the
impulse approximation model, one finds that $A_{NN}=-0.26\pm0.02$.
This is to be compared to the SAID value of $-0.28$, though no error
can be deduced directly on their prediction~\cite{ARN00}. All this
suggests that the methodology applied by the ANKE collaboration is
sufficient to deliver useful $np$ amplitudes at higher energies,
where less is known experimentally. Compared to the SPES~IV and EMRIC
experiments, there are finer divisions in momentum transfer and hence
more points in the safe $q$ region.

Apart from taking data up to the maximum COSY energy of $T_d\approx
2.3$~GeV, there are plans to measure the deuteron charge exchange
reaction with a polarised beam and target~\cite{SPI05}. The resulting
values of the two transverse spin correlation parameters will allow
the relative phases of the spin-flip amplitudes to be determined.

To go higher in energy, it will be necessary to use a proton beam on
a deuterium target, detecting both slow recoil protons from the
$p\vec{d}\to \{pp\}n$ in the silicon tracking telescopes with which
ANKE is equipped~\cite{SCH03}. The drawback here is that the
telescopes require a minimum momentum transfer so that the energies
of the protons can be measured and this is of the order of
150~MeV/$c$ at low $E_{pp}$. This technique has already been used at
CELSIUS to generate a tagged neutron beam on the basis of the $pd\to
npp$ reaction at 200~MeV by measuring both slow recoil protons in
silicon microstrip detectors~\cite{PET04}.
%
%
\subsection{Data summary}

\begin{figure}[hbt]
\begin{center}
\includegraphics[width=0.7\columnwidth,clip]{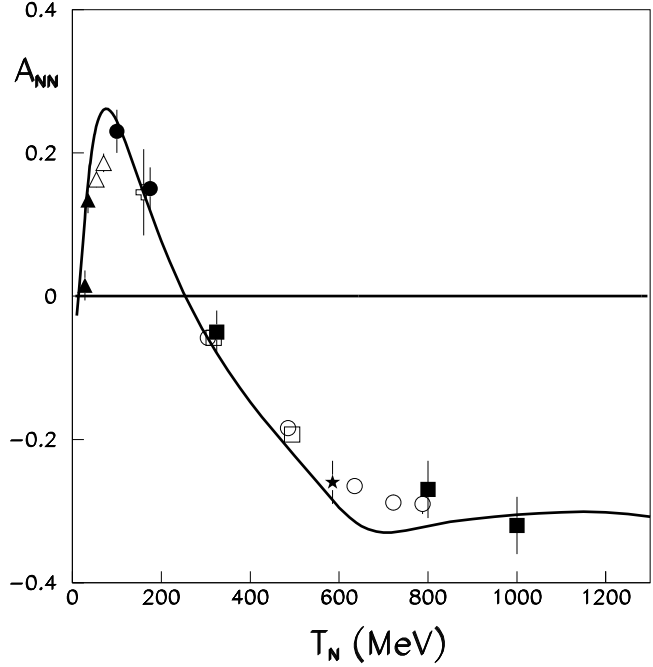}
\caption{Values of the forward deuteron tensor analysing power in the
$\vec{d}p\to\{pp\}n$ reaction as a function of the kinetic energy per
nucleon $T_N$. The directly measured experimental data (closed
symbols) from SPES~IV (squares)~\cite{SAM95}, EMRIC (closed
circles)~\cite{KOX93}, ANKE (star)~\cite{CHI09}, and RCNP
(triangles)~\cite{MOT88} were all obtained with a $pp$ excitation
energy of 1~MeV or less. The error bars include some estimate of the
uncertainty in the extrapolation to $\theta=0$. The open symbols were
obtained from measurements of the polarisation transfer parameter in
$d(\vec{p},\vec{n})pp$ by using Eq.~\eqref{Ohlsen}. The data are from
Refs.~\cite{MCN92} (circles), \cite{MER93} (squares), \cite{SAK87}
(cross), and \cite{PIC90,ZEI99} (triangles). The curve is the plane
wave $^{1\!}S_0$ prediction of Eq.~\eqref{ann_pred}, as tabulated in
Table~\ref{table1}. \label{annsum}}
\end{center}
\end{figure}

In Table~\ref{Ayy} we present the experimental values of the deuteron
tensor analysing power in the $\vec{d}p\to\{pp\}n$ reaction
extrapolated to the forward direction. The error bars include some
attempt to take into account the uncertainty in the angular
extrapolation. The resulting data are also shown in
Fig.~\ref{annsum}.

\begin{table}[thb]
\caption{Measured values of the forward deuteron tensor analysing
power $A_{NN}$ in the $\vec{d}p\to\{pp\}n$ reaction in terms of the
kinetic energy per nucleon $T_N$. The errors include some estimate
for the extrapolation to $\theta=0^{\circ}$.\label{Ayy}}
\begin{center}
\begin{tabular}{|c|c|c|c|c|}
\hline
 &&&& \\
 $T_{N}$ & $A_{NN}(0)$       & Facility    & Year & Ref.  \\
  (MeV)    &                   &               &      &       \\
\hline
  ~~28 & $0.015\pm0.021$ & RCNP   & 1987   &\cite{MOT88}  \\
\hline
  ~~35 & $0.134\pm0.018$ & RCNP   & 1987   &\cite{MOT88}  \\
\hline
  ~100 & $0.23\pm0.03$   & EMRIC   & 1993   &\cite{KOX93}  \\
\hline
  ~175 & $0.15\pm0.03$   & EMRIC   & 1993   &\cite{KOX93}  \\
\hline
  ~325 & $-0.05\pm0.03$ & SPES~IV   & 1995   &\cite{SAM95}  \\
\hline
  ~585 & $-0.26\pm0.03$ & ANKE   & 2009   &\cite{CHI09}  \\
\hline
  ~800 & $-0.27\pm0.04$ & SPES~IV   & 1995   &\cite{SAM95}  \\
\hline
  1000 & $-0.32\pm0.04$ & SPES~IV   & 1995   &\cite{SAM95}  \\
\hline
\end{tabular}
\end{center}
\end{table}

In the forward direction the plane wave impulse approximation
predictions of Eq.~\eqref{ann_pred} for the forward analysing power
should be quite accurate provided that the excitation energy in the
final diproton is small so that it is in the $^{1\!}S_0$ state. This
condition is well met by the data described here, where $E_{pp}$ is
always below 1~MeV~\cite{SAM95,MOT88,KOX93,CHI09}. This prediction,
which is also tabulated in Table~\ref{table1}, describes the trends
of the data very well in regions where the neutron-proton phase
shifts are well determined.

We also show in the figure the values of $A_{NN}$ deduced using
Eq.~\eqref{Ohlsen} from the $d(\vec{p},\vec{n})pp$ measurements
summarised in Table~\ref{table_transfer}. Only those data are
retained where the neutron polarisation was well measured and the
$pp$ excitation energy was small, though generally not as well
determined as when the two final protons were detected. The
consistency between the $(\vec{d},pp)$ and $(\vec{p},\vec{n})$ data
is striking and it is interesting to note that they both suggest
values of $A_{NN}$ that are slightly lower in magnitude at high
energies than those predicted by the $np$ phase shifts of the SAID
group~\cite{ARN00}. The challenge now is to continue measuring these
data into the more unchartered waters of even higher energies.

Although we have concentrated here on the results for the forward
analysing power, it is clear that this represents only a small part
of the total data set as demonstrated by the results of
Figs.~\ref{Sams}, \ref{Serge}, and \ref{David}.

\clearpage
%
%
\section{Conclusions}
\label{conclusions}

Originally the deuteron was thought of merely as a useful substitute
for a free neutron target. As an example of this, it has been shown
that at large momentum transfers the spin-dependent parameters
measured in free $np$ scattering and quasi-free in $pd$ collisions
give very similar results~\cite{ADL99}. The situation is very
different at low momentum transfers where it is not clear which of
the nucleons is the spectator or, indeed, whether the concept of
calling one of the nucleons a spectator makes any sense at all.
However, a more interesting effect comes about in the medium energy
neutron charge exchange on the deuteron, $nd \to p\{nn\}$, when the
excitation energy $E_{nn}$ in the two neutron system is very low.
Under such conditions the Pauli principle demands that the two
neutrons should be in a $^{1\!}S_0$ state and there then has to be
spin-flip isospin-flip transition from the spin-triplet $np$ in the
deuteron to the singlet $nn$ system. The rate for the charge-exchange
deuteron breakup $nd \to p\{nn\}$ would then depend primarily on the
spin-spin $np\to pn$ amplitudes.

The above remarks only assume a practical importance because of an
``accident'' in the low energy nucleon-nucleon interaction. In the
$nn$ system there is an antibound (or virtual) state pole only a
fraction of an MeV below threshold. Although the pole position is
displaced slightly in the $pp$ case by the Coulomb repulsion, it
results in huge $pp$ and $nn$ scattering lengths. In the $nd \to pnn$
reaction, it leads to the very characteristic peak at the hard end of
the momentum spectrum of the produced proton. Since we know that
these events are the result of the spin-flip interaction, we clearly
want to use them to investigate in greater depth this interaction.
There are two distinct ways to try to achieve our aims and we have
tried to review them both in this article. These are the inclusive
(sum-rule) approach of section~\ref{QECE} and the high resolution
polarisation experiments of sections~\ref{transfer} and \ref{polar}.

In impulse approximation, at zero momentum transfer, the $d(n,p)nn$
interaction only excites spin-singlet final states and
Dean~\cite{DEA72,DEA72A} has shown that the inclusive measurement of
the proton momentum spectrum can then be interpreted in terms of the
spin-flip $np$ amplitudes through the use of a sum rule. Though the
shape of the proton momentum spectrum must depend upon the details of
the low energy $nn$ interaction and also on the deuteron $D$-state,
the integral over all momenta would not, provided that the sum rule
has converged before any of the limitations imposed by the three-body
phase space have kicked in.

The inclusive approach has many positive advantages, in addition to
being independent of the low energy nucleon-nucleon dynamics. In a
direct comparison of the production rates of protons in the
$d(n,p)nn$ and $p(n,p)n$ reactions using the same apparatus, many of
the sources of systematic errors drop out in the evaluation of the
cross section ratio $R_{np}(0)$. These are primarily effects
associated with the neutron flux and uncertainties in the proton
detection system.

There are, however, no similar benefits when working with a proton
beam, where one measures instead $d(p,n)pp$. Here one can only
construct the $R_{pn}(0)$ ratio by dividing by a $p(n,p)n$ cross
section that has been measured in an independent experiment. This is
probably the reason why there are fewer entries in Table~\ref{table3}
compared to Table~\ref{table2}. We must therefore stress that, in
general, the $d(np)nn$ determinations of $R_{np}(0)$ are much to be
preferred over those of $d(p,n)pp$.

On the face of it, the determination of $R_{pn}(0)$ through the
measurement of the two fast protons from the $p(d,pp)n$ reaction in a
bubble chamber looks like a very hard way to obtain a
result~\cite{GLA08}. In addition to having to use independent data to
provide the normalisation cross section in the denominator, the
reaction is first measured exclusively in order afterwards to
construct an inclusive distribution. On the other hand, a full
kinematic determination allows one to check many of the assumptions
made in the analysis and, in particular, those related to the
isolation of the charge-exchange impulse approximation contribution
from those of other possible Feynman diagrams.

A major difficulty in any of the inclusive measurements is ensuring
that the phase space is sufficiently large that the sum rule has been
saturated without being contaminated by other driving mechanisms.
This means that the low energy determinations of $R_{np}(0)$ are all
likely to underestimate the ``true'' value and there could be some
effects from this even through the energy range of the PSI
experiments~\cite{PAG88}. Even more worrying is the fact that at low
energies the rapid variation of $K_{NN}(0)$ with $E_{pp}$, as
measured in the $d(\vec{p},\vec{n})pp$ reaction~\cite{ZEI99}, shows
that there are significant deviations from plane wave impulse
approximation with increasing $E_{pp}$. These deviations are probably
too large to be ascribed to effects arising from the variation of the
longitudinal momentum transfer with $E_{pp}$. This brings into
question the whole sum rule approach at low energies.

The alternative high resolution approach of measuring the $^{1\!}S_0$
peak of the final state interaction requires precisely that,
\textit{i.e.}, high resolution. This can be achieved in practice by
measuring the $(n,p)$ reaction with a very long time-of-flight
path~\cite{MER93} or by measuring the protons in the $dp\to\{pp\}n$
reaction with either a deuteron beam~\cite{SAM95,KOX93,CHI09} or a
very low density deuterium target~\cite{EST65}. The resulting data
are then sensitive to the low energy $np$ interaction in the deuteron
and the $pp$ interaction in the $^{1\!}S_0$ final state. However,
such interactions are well understood and lead to few ambiguities in
the charge exchange predictions. Establishing a good overall
normalisation can present more of a challenge. In addition to obvious
acceptance and efficiency uncertainties, if one evaluates a cross
section integrated up to say $E_{pp}=3$~MeV then one has to measure
the 3~MeV with good absolute precision, which is non-trivial for a
deuteron beam in the GeV range. Hence it might be that at high
energies the inclusive measurements could yield more precise
determinations of absolute values of $R_{np}(0)$~\cite{SHA09} than
could be achieved by using high resolution experiments.

On the other hand, measuring just the FSI peak with good resolution
allows one more easily to follow the variation with momentum transfer
and there are also fewer kinematic ambiguities. More crucially, the
spin information from the $(\vec{n},\vec{p}\,)$ or $(\vec{d},\{pp\})$
reactions enables one to separate the different spin contributions to
the small angle charge exchange cross section. It could of course be
argued that this is not just a benefit for an exclusive reaction
since, if the Dubna bubble chamber experiments~\cite{GLA08} had been
carried out with a polarised deuteron beam, then these would also
have been able to separate the contributions from the two independent
forward spin-spin contributions through the use of the generalised
Dean sum rule~\cite{BUG87,LEH08}. It is, however, much more feasible
to carry out $(d,\{pp\})$ measurements with modern electronic
equipment and the hope is that, through the use of polarised beams
and targets, they will lead to evaluations of the relative phases
between the three independent $np\to pn$ spin-spin amplitudes out to
at least $q\approx m_{\pi}$~\cite{SPI05}.

We have been very selective in this review, concentrating our
attention on the forward values of the $nd\to pnn/np\to pn$ cross
section ratio, the $(\vec{n},\vec{p}\,)$ polarisation transfer, and
the deuteron tensor analysing power in nucleon-deuteron
charge-exchange break-up collisions. In the latter cases, we have
specialised to the kinematic situations where two of the final
nucleons emerge in the $^{1\!}S_0$ state. Under these conditions
there are strong connections between the three types of experiment
described and this we have tried to stress. However, there is clearly
much additional information in the data at larger angles, which we
have here generally neglected. We have also avoided discussing the
extensive data that have been taken on nuclear targets, where the
selectivity of the $(\vec{n},\vec{p}\,)$ or $(\vec{d},\{pp\})$
reactions can be used to identify particular classes of final nuclear
states. At the higher energies, these states could even include the
excitation of the $\Delta(1232)$ isobar.

Despite the successful measurements, none of the $R_{np}(0)$ data nor
those from the exclusive polarised measurements have so far been
included in any of the existing phase shift analyses. They have
merely been used as \textit{a posteriori} checks on their
predictions. We have argued that they could also provide valuable
input into the direct neutron-proton amplitude reconstruction in the
backward direction~\cite{LEH08}. For any of these purposes it would
be highly desirable to control further the range of validity of the
models used to interpret the data and, in particular, to examine
further the effects of multiple scattering. There remain therefore
theoretical as well as experimental challenges to be overcome.
%
%
\section*{Acknowledgements}
We are grateful to J.~Ludwig for furnishing us with a partial copy of
Ref.~\cite{PAG88}. Several people gave us further details regarding
their own experiments. These include D.~Chiladze, M.J.~Esten,
V.~Glagolev, A.~Kacharava, S.~Kox, M.W.~McNaughton, T.~Motobayashi,
I.~Sick, and C.A.~Whitten. There were also helpful discussions with
D.V.~Bugg, Z.~Janout, N.~Ladygina, I.I.~Strakovsky, E.~Strokovsky,
and Yu.~Uzikov. One of the authors (CW) wishes to thank the Institute
of Experimental and Applied Physics of the Czech Technical University
Prague, and its director Stanislav Posp\'i\v{s}il, for hospitality
and support during the preparation of this paper. This work has been
supported by the research programme MSM~684~077~0029 of the Ministry
of Education, Youth, and Sport of the Czech Republic.

%
%

\end{document}